\documentclass[showpacs,showkeys,aps,prd,twocolumn,floatfix]{revtex4-2}

\usepackage[pagewise,mathlines,displaymath]{lineno}
\usepackage{amsmath}
\usepackage{amssymb}

\usepackage{CJK,upgreek,fancyhdr}
\usepackage{tikz}
\usepackage{graphicx}
\usepackage{float}
\usepackage[caption = false]{subfig}
\usepackage{dcolumn}
\usepackage{bm}
\usepackage{gensymb}

\usepackage{tabstackengine}
\setstackEOL{\cr}
\setstackgap{L}{\normalbaselineskip}

\newcolumntype{L}[1]{>{\raggedright\arraybackslash}p{#1}}
\newcolumntype{C}[1]{>{\centering\arraybackslash}p{#1}}
\newcolumntype{R}[1]{>{\raggedleft\arraybackslash}p{#1}}
\newcommand\T{\rule{0pt}{2.6ex}}       
\newcommand\B{\rule[-1.2ex]{0pt}{0pt}} 

\usepackage{hyperref}
\hypersetup{
 pdftitle={},%
 pdfauthor={},%
 pdfsubject={},%
 pdfkeywords={},%
 pdfstartview={},%
 bookmarksopen=true, breaklinks=true, debug=true,
 colorlinks=true, linkcolor=blue, citecolor=red, urlcolor=blue,
 hyperfigures=true
}

\def \b{{\cal B}}

\begin{document}
\begin{CJK*}{UTF8}{gkai}

\title{\boldmath Measurement of the absolute branching fractions for purely leptonic $D_s^+$ decays}

\author{
\begin{small}
\begin{center}
M.~Ablikim(麦迪娜)$^{1}$, M.~N.~Achasov$^{10,c}$, P.~Adlarson$^{67}$, S. ~Ahmed$^{15}$, M.~Albrecht$^{4}$, R.~Aliberti$^{28}$, A.~Amoroso$^{66A,66C}$, M.~R.~An(安美儒)$^{32}$, Q.~An(安琪)$^{49,63}$, X.~H.~Bai(白旭红)$^{57}$
% ...
\\
\vspace{0.2cm}
(BESIII Collaboration)\\
\vspace{0.2cm} {\it
$^{1}$ Institute of High Energy Physics, Beijing 100049, People's Republic of China\\
% ...
$^{4}$ Bochum Ruhr-University, D-44780 Bochum, Germany\\
% ...
$^{10}$ G.I. Budker Institute of Nuclear Physics SB RAS (BINP), Novosibirsk 630090, Russia\\
% ...
$^{15}$ Helmholtz Institute Mainz, Johann-Joachim-Becher-Weg 45, D-55099 Mainz, Germany\\
% ...
$^{28}$ Johannes Gutenberg University of Mainz, Johann-Joachim-Becher-Weg 45, D-55099 Mainz, Germany\\
% ...
$^{32}$ Liaoning Normal University, Dalian 116029, People's Republic of China\\
% ...
$^{49}$ State Key Laboratory of Particle Detection and Electronics, Beijing 100049, Hefei 230026, People's Republic of China\\
% ...
$^{57}$ University of Jinan, Jinan 250022, People's Republic of China\\
% ...
$^{63}$ University of Science and Technology of China, Hefei 230026, People's Republic of China\\
% ...
$^{66}$ (A)University of Turin, I-10125, Turin, Italy; (B)University of Eastern Piedmont, I-15121, Alessandria, Italy; (C)INFN, I-10125, Turin, Italy\\
$^{67}$ Uppsala University, Box 516, SE-75120 Uppsala, Sweden\\
% ...
$^{c}$ Also at the Novosibirsk State University, Novosibirsk, 630090, Russia\\
% ...
}\end{center}
\vspace{0.4cm}
\end{small}}

\date{\today}

\begin{abstract}
 We report new measurements of the branching fraction $\b(D_s^+\to \ell^+\nu)$, where $\ell^+$ is either $\mu^+$
or $\tau^+(\to\pi^+\bar{\nu}_\tau)$, based on $6.32$ fb$^{-1}$ of electron-positron annihilation data 
collected by the BESIII experiment at six center-of-mass energy points between $4.178$ and $4.226$~GeV.
Simultaneously floating the $D_s^+\to\mu^+\nu_\mu$ and $D_s^+\to\tau^+\nu_\tau$ components
yields $\b(D_s^+\to \tau^+\nu_\tau) = (5.21\pm0.25\pm0.17)\times10^{-2}$,
$\b(D_s^+\to \mu^+\nu_\mu) = (5.35\pm0.13\pm0.16)\times10^{-3}$, and
the ratio of decay widths
$R=\frac{\Gamma(D_s^+\to \tau^+\nu_\tau)}{\Gamma(D_s^+\to \mu^+\nu_\mu)} = 9.73^{+0.61}_{-0.58}\pm 0.36$,
where the first uncertainties are statistical and the second systematic.
No evidence of {\it CP} asymmetry is observed in the decay rates $D_s^\pm\to\mu^\pm\nu_\mu$ 
and  $D_s^\pm\to\tau^\pm\nu_\tau$: $A_{\it CP}(\mu^\pm\nu) = (-1.2\pm2.5\pm1.0)\%$
and $A_{\it CP}(\tau^\pm\nu) = (+2.9\pm4.8\pm1.0)\%$.  Constraining our measurement to the Standard Model
expectation of lepton universality ($R=9.75$), we find the more precise results
$\b(D_s^+\to \tau^+\nu_\tau) = (5.22\pm0.10\pm 0.14)\times10^{-2}$ and
$A_{\it CP}(\tau^\pm\nu_\tau) = (-0.1\pm1.9\pm1.0)\%$.
Combining our results with inputs external to our analysis, we
determine
the $c\to \bar{s}$ quark mixing matrix element, $D_s^+$ decay constant, and ratio of the decay constants to be
$|V_{cs}| =  0.973\pm0.009\pm0.014$, 
$f_{D^+_s} = 249.9\pm2.4\pm3.5~\text{MeV}$, 
and $f_{D^+_s}/f_{D^+} = 1.232\pm0.035$, respectively.
\end{abstract}

\maketitle
\end{CJK*}

\section{INTRODUCTION}\label{sec:intro}

Purely leptonic decays of heavy mesons are the subject of great
experimental and theoretical interest
because of their potential for precise tests of the Standard Model
(SM), including determination of
the Cabibbo-Kobayashi-Maskawa (CKM) matrix elements and sensitivity to non-SM physics.
Leptonic decays of charmed mesons play an important role in this, with clean experimental signatures
and the opportunity for rigorous tests of strong-interaction theory, especially lattice-QCD (LQCD) calculations.
In the decay process $D_s^+ \rightarrow \ell^+ \nu_\ell$, the charm quark ($c$) and antistrange quark ($\bar{s}$)
annihilate through a virtual $W$ boson to a
charged and neutral lepton pair.
(Throughout this article, charge conjugate modes
are implied unless otherwise noted.)
According to the SM,
the branching fraction for this process (ignoring radiative corrections) is
given as follows:
\begin{equation}\label{eq:decayrate}
  \b(D^+_s\to\ell^+\nu_\ell) = \tau_{D_s^+}\frac{G^2_F}{8\pi}f^2_{D^+_s}m^2_\ell m_{D^+_s}\left(1-\frac{m^2_\ell}{m^2_{D^+_s}}\right)^2|V_{cs}|^2,
\end{equation}
\noindent where $\tau_{D_s^+}$ is the $D_s^+$ lifetime, $m_{D^+_s}$ is the mass of $D_s^+$, $m_\ell$ is the mass of
the charged lepton ($e^+$, $\mu^+$, or $\tau^+$), and
$G_F$ is the Fermi coupling constant, all of which are known
to precision~\cite{pdg2020}. The remaining two factors, $|V_{cs}|$ and $f_{D^+_s}$, must be determined experimentally and 
are of great interest.
(1) $V_{cs}$ is a fundamental SM parameter, the CKM matrix element describing the coupling between the $c$ and $\bar{s}$ quarks. (2) $f_{D^+_s}$ is the $D_s^+$ decay constant, the amplitude for  quark-antiquark annihilation inside the meson, which can be thought of as the overlap of the wave functions of 
$c$ and $\bar{s}$ at zero spatial separation.

It follows from Eq.~\eqref{eq:decayrate} that measurement of the branching fraction $\b(D^+_s\to\ell^+\nu_\ell)$ is 
essentially a determination of $f^2_{D^+_s}|V_{cs}|^2$.  In practice, we can
determine
$f_{D^+_s}$ by combining a 
measurement of $\b(D^+_s\to\ell^+\nu_\ell)$ with an independent determination of $|V_{cs}|$, thereby testing
theoretical predictions, primarily from LQCD.  Testing the LQCD calculations in $D$ and $D_s^+$ decays
is especially important to validate their application to the $B$-meson sector, in which the precision
of experimental determination of $f_{B^+}$ is very limited due to the small value of 
$|V_{ub}|$, $0.00361^{+0.00011}_{-0.00009}$~\cite{pdg2020}.  It is also possible to reverse this approach, 
determining
$|V_{cs}|$ from $\b(D^+_s\to\ell^+\nu_\ell)$ with a theoretical estimate of $f_{D^+_s}$ from LQCD 
and comparing the result with other experimental determinations of $|V_{cs}|$.

The ratio of decay widths for leptonic decays to $\mu$ and $\tau$ is also an interesting quantity to measure:
\begin{equation}\label{eq:theR}
  R = \frac{\Gamma(D^+_s\to\tau^+\nu_\tau)}{\Gamma(D^+_s\to\mu^+\nu_\mu)}
  = \frac{m^2_\tau(1-\frac{m^2_\tau}{m^2_{D_s}})^2}{m^2_\mu(1-\frac{m^2_\mu}{m^2_{D_s}})^2}.
\end{equation}
\noindent In this ratio, the decay constant and the CKM element cancel, giving a very precise SM prediction of $R = 9.75\pm0.01$.
Any deviation from this value potentially indicates
the existence of non-SM physics.

Using the known values of the lepton masses and other constants~\cite{pdg2020}, the measured  
$D^+_s$ lifetime~\cite{pdg2020} (including recent improvements in precision by the LHCb 
Collaboration~\cite{Dslifetime}), the weighted average of
recent four-flavor LQCD calculations~\cite{flag2019} ($f_{D^+_s} = 249.9\pm0.5$~MeV),
and the latest determination of the $c\to\bar{s}$ coupling from the
global fit of CKM parameters~\cite{pdg2020} ($|V_{cs}| = 0.97320\pm0.00011$), one arrives at the 
following SM predictions of the $D_s^+$ leptonic branching fractions:
$\b(D^+_s\to e^+\nu_e) = (1.261\pm0.004)\times10^{-7}$,
$\b(D^+_s\to\mu^+\nu_\mu) = (5.360\pm0.017)\times10^{-3}$, and
$\b(D^+_s\to\tau^+\nu_\tau) = (5.221\pm0.018)\%$.
In this article, we report new
measurements of the branching fractions
$\b(D^+_s\to\mu^+\nu_\mu)$
and $\b(D^+_s\to\tau^+\nu_\tau)$ (via $\tau^+\to\pi^+\bar{\nu}_\tau$),
and of the {\it CP}-violating asymmetries $A_{\it CP}(\ell^\pm\nu_\ell)$.
These measurements have been made with
$6.32$ fb$^{-1}$ of $e^+e^-$ annihilation data
collected
at center-of-mass energies between $4178$ and $4226$~MeV
with the BESIII detector~\cite{2009MAblikimDet}
at the Beijing Electron Positron Collider (BEPCII)~\cite{bepcii}.
This work, which uses a  larger data sample and a procedure that is simultaneously
sensitive to both $D^+_s\to\mu^+\nu_\mu$ and $D^+_s\to\tau^+\nu_\tau$ decays, is
distinct from and supersedes our previous measurement
of $\b(D^+_s\to\mu^+\nu_\mu)$~\cite{bes3dsmunu}.

\section{THE BESIII EXPERIMENT AND DATA SETS}

BESIII~\cite{2009MAblikimDet} is a cylindrical spectrometer
with a geometrical acceptance 
of $93\%$ of $4\pi$.
It consists of a small-celled, helium-based main drift chamber (MDC),
a plastic scintillator time-of-flight system (TOF),
a CsI(Tl) electromagnetic calorimeter (EMC),
a superconducting solenoid providing a $1.0$-T magnetic field, and
a muon counter (MUC).
The charged particle momentum resolution is $0.5\%$ at a transverse momentum of 1 ${\rm GeV}/c$.
The specific ionization ($dE/dx)$ measurement provided by the MDC
has a resolution of $6\%$, and
provides $3\sigma$ separation of
  charged pions and kaons.
The time resolution of the TOF is $80$~ps ($110$~ps) in the central barrel (end-cap) region of the
detector.  The end-cap TOF system was upgraded with multigap resistive plate chamber technology in 
$2015$~\cite{bes3mrpc}, improving its time resolution to $60$~ps.
Approximately $83\%$ of the sample employed
  in this work was taken with the improved configuration.
The energy resolution for photons is $2.5\%$ ($5\%$) at $1$~GeV in the barrel (end-cap) region of the EMC. 
A more detailed description of the BESIII detector is given in Ref.~\cite{2009MAblikimDet}.

The data samples employed in this work were taken at six $e^+e^-$ center-of-mass energies 
($E_{\text{cm}}$ = $4178$, $4189$, $4199$, $4209$, $4219$, and around $4226$~MeV~\cite{ecm4230}).  
The integrated luminosities for these subsamples are
$3189.0$, $526.7$,  $526.0$, $517.1$, $514.6$, and $1047.3$~pb$^{-1}$~\cite{lumi4230}, respectively.
For simplicity, we refer to these datasets as $4180$, $4190$, $4200$, $4210$, $4220$, and $4230$ in 
the rest of this article.  For some aspects of our analysis, especially in assessing systematic 
uncertainties, we organize the samples into three groups that were acquired during the same year under 
consistent running conditions.   The $4180$ sample was taken in 2016, while the group $4190$-$4220$ was taken 
in 2017 and $4230$ was taken in 2013. 

To assess background processes and determine detection efficiencies, we produce and analyze 
{\scshape{geant4}}-based~\cite{geant} Monte Carlo (MC) simulation samples for all six datasets,
with sizes that are $40$ times the integrated luminosity of data (``$40\times$'').
The MC samples are produced 
using known decay rates~\cite{pdg2020} and correct angular distributions by two event generators,
 {\sc EvtGen}~\cite{evtgen} for charm and charmonium decays and {\sc KKMC}~\cite{kkmc} for continuum processes.
The samples consist of $e^+e^-\to D\bar{D}$, $D^*D$, $D^*D^*$, $D_sD_s$, 
$D_s^*D_s$, $D_s^*D_s^*$, $DD^*\pi$, $DD\pi$, $q\bar{q} (q = u, d, s)$, $\gamma J/\psi$, $\gamma \psi(3686)$, and $\tau^+\tau^-$.  Charmonium decays that are not accounted for by exclusive measurements are simulated by  {\sc Lundcharm}~\cite{lundcharm}.
Additionally we generate separate samples consisting only of signal events,
with size $300$ times larger than is expected in our data.
All MC simulations include the effects of initial-state radiation (ISR) and final-state radiation (FSR).  We simulate ISR with {\sc ConExc}~\cite{conexc}
for $e^+e^- \to c\bar{c}$ events within the framework of EvtGen, and with KKMC for noncharm continuum processes.
FSR is simulated with PHOTOS~\cite{photos}.

\section{Analysis method}

We employ the double-tag technique pioneered by the MARK III Collaboration~\cite{markiii} in our selection 
of $D_s^\pm\to\ell^\pm\nu_\ell$ decays in $e^+e^-\to D^{*\pm}_sD^\mp_s$ events.  In this method, a 
$D^-_s$ is fully reconstructed through one of several hadronic decay modes (tag side), while 
we reconstruct only one charged track from the $D^+_s$ 
(signal side).
Note that the reconstructed tag-side $D_s$ can either be directly produced
from the $e^+e^-$ collision (direct) or be the daughter of a $D_s^*$ (indirect).
We reconstruct the radiative photon from $D_s^*\to\gamma D_s$ in the tag-side 
selection and use it in the analysis of the signal side.  (Further explanation is provided in 
Sec.~\ref{sec:STselect}.) The absolute branching fraction determined by this method
does not depend on the integrated luminosity or the produced number of $D^{*\pm}_sD^\mp_s$ pairs:
\begin{equation}\label{eq:dteq}
\b(D^+_s\to\mu^+\nu_\mu) = \frac{N_{DT}^{\mu\nu}}{\sum_{i}{N^i_{ST}(\epsilon^{\mu\nu,i}_{DT}/\epsilon^i_{ST})}},
\end{equation}
\noindent where $N_{DT} = \sum_{i}{N^i_{DT}}$ is the
summed yield over tag modes of double-tag ($DT$) events in which the tag side and signal side are simultaneously reconstructed,
$N_{ST}^i$ is the single-tag ($ST$) yield of reconstructed $D^-_s$ for tag mode $i$, and
 $\epsilon^{\mu\nu,i}_{DT}$ and $\epsilon^i_{ST}$ are the corresponding
reconstruction efficiencies.
Similarly for  $\b(D^+_s\to\tau^+\nu_\tau)$ we have
\begin{equation}\label{eq:dteqtau}
\b(D^+_s\to\tau^+\nu_\mu) = \frac{N_{DT}^{\tau\nu}}{\sum_{i}{N^i_{ST}(\epsilon^{\tau\nu,i}_{DT}/\epsilon^i_{ST})} \b(\tau^+\to\pi^+\bar{\nu}_\tau)},
\end{equation}
\noindent where $\epsilon^{\tau\nu,i}_{DT}$ does not include
$\b(\tau^+\to\pi^+\bar{\nu}_\tau)$
but $\epsilon^{\ell\nu,i}_{DT}$ does include $\b(D^*_s\to\gamma D_s)$.

In the ratios of $N_{DT}^{\ell\nu}/N^i_{ST}$ and $\epsilon^{\ell\nu,i}_{DT}/\epsilon^i_{ST}$, systematic
uncertainty associated with the tag-side analysis
mostly cancels, except for
a possible uncertainty due to variations in $ST$
reconstruction efficiencies (tag bias, discussed
in Sec.~\ref{sec:systnonfit}).

\subsection{\boldmath  Selection of tagged $D_s^-$ candidates}\label{sec:STselect}

The $ST$ $D^-_s$ is reconstructed using tracks and EMC showers
that pass several quality requirements.
The selection criteria for $D^-_s$ daughters and the reconstruction procedures are
the same as those described in Ref.~\cite{ddpp}.
Tracks must be within the fiducial region  ($|\cos\theta| < 0.93$, where $\theta$ is the polar angle
relative to the positron beam direction)
and originate within $1$~cm ($10$~cm) of the interaction point
in the plane transverse to the beam direction (along the beam direction).
This requirement on the primary vertex is not applied  for the reconstruction of $K^0_S\to\pi^+\pi^-$,
for which we constrain the charged pion pair to have a common vertex with a loose fit-quality requirement of
$\chi^2 < 100$ for $1$ degree of freedom.
To be selected as a photon candidate, an EMC shower must 
not be associated with any charged track~\cite{Ablikim:2015ycp},
must have an EMC hit time between $0$ and $700$~ns
to suppress activity that is not consistent with originating from the collision event,
and must have an energy of
at least $25$~MeV if it is in the barrel region of the detector ($|\cos\theta| < 0.8$) 
and $50$~MeV if it is in the end-cap region ($0.86 < |\cos\theta| < 0.92$)
to suppress noise in the EMC.

We apply $K/\pi$ particle identification (PID) based on TOF and $dE/dx$ measurements,
with the identity as a pion or kaon assigned based on which 
hypothesis has the higher likelihood.  Pions from the intermediate states $K^0_S\to\pi^+\pi^-$, 
$\eta\to\pi^+\pi^-\pi^0$ and $\eta'\to\pi^+\pi^-\eta$ are not required to satisfy the 
$K/\pi$-identification requirement.
We also require that the reconstructed momentum for any charged or
neutral pion have a magnitude of at least $100$~MeV$/c$ to suppress
events from $D^*\to D\pi$.

We select tag modes for this analysis to maximize signal sensitivity 
and minimize tag bias by performing the entire analysis procedure
  on our cocktail MC sample with various combinations of tag modes.
The following thirteen $D_s^-$ hadronic decay modes are used:
$D^-_s\to K^0_SK^-$, $K^-K^+\pi^-$, $K^0_SK^-\pi^0$,
$K^-K^+\pi^-\pi^0$, $K^0_SK^-\pi^+\pi^-$, $K^0_SK^+\pi^-\pi^-$,
$\pi^-\pi^-\pi^+$, $\pi^-\eta$,
$\rho^-\eta$, $\rho^-\eta_{3\pi}$,
$\pi^-\eta'_{\pi\pi\eta}$, $\pi^-\eta'_{\gamma\rho}$, and
$K^-\pi^-\pi^+$, where
$\pi^0\to\gamma\gamma$,
$K^0_S\to\pi^+\pi^-$,
$\eta\to\gamma\gamma$,
$\eta_{3\pi}\to\pi^+\pi^-\pi^0$,
$\rho^{-(0)}\to\pi^{0(+)}\pi^-$,
$\eta'_{\pi\pi\eta}\to\pi^+\pi^-\eta$, and
$\eta'_{\gamma\rho}\to\gamma\rho^0$.
Requirements on invariant-mass ranges for the intermediate states
$\pi^0$, $K^0_S$, $\eta$, $\eta_{3\pi}$, $\rho$, $\eta'_{\pi\pi\eta}$, and $\eta'_{\gamma\rho}$
are chosen to cover
$\pm(3-4)\sigma$
of the signal mass resolutions, with the exception of $\rho$,
for which $520~(500) < M_{\pi\pi} < 1000$~MeV$/c^2$ is selected for the charged (neutral) case.
We also require an isolation criterion for photons used in the reconstruction of
$\eta'\to\gamma\rho^0$; the reconstructed photons must be separated from the extrapolated
positions of all charged particles by more than $20\degree$.
For the decay mode $D^-_s\to K^-\pi^-\pi^+$ we exclude the dipion mass range
$486< M_{\pi\pi}< 510$~MeV$/c^2$ to avoid overlap with the
 $D^-_s\to K^0_SK^-$ mode.

\begin{figure}[htbp]
\centering
\includegraphics[keepaspectratio=true,width=3.1in,angle=0]{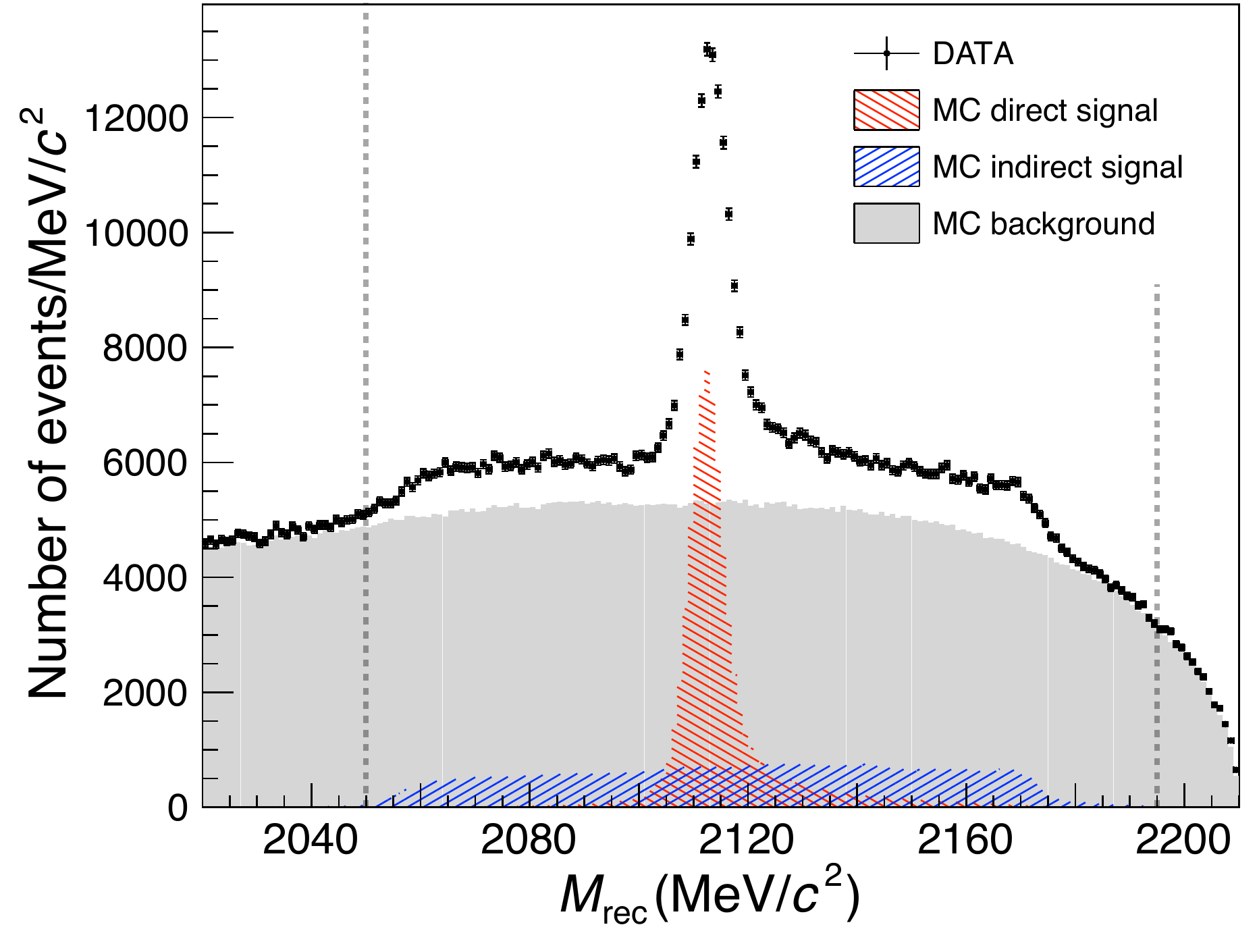}
\caption{Recoil mass distributions against $D_s^-\to K^-K^+\pi^-$ for $4180$ data and 
MC-simulated background, which mostly consists of charm decays and continuum processes.}
\label{fig:strecoil401}
\end{figure}

 Once the tag-side $D_s^-$ is reconstructed, we preselect our sample by requiring that the 
invariant mass of the reconstructed $D_s^-$ satisfy the condition
$1900<M_{\text{inv}}(D^-_s)<2030$~MeV$/c^2$.
Then we calculate the recoil mass against the $D^-_s$ tag inclusively as
$M_{\text{rec}}^2c^4 = \Big (E_{\text{cm}} - \sqrt{|\vec{p}_{D^-_s}|^2c^2+m^2_{D_s}c^4}\Big )^2 -
|\vec{p}_{D^-_s}|^2c^2$
in the center-of-mass system of the initial
$e^+e^-$, where $E_{\text{cm}} = 2\times E_{\text{beam}}$, $\vec{p}_{D^-_s}$ is the three-momentum
of the reconstructed $D^-_s$, and $m_{D_s}$ is the known $D_s$ mass~\cite{pdg2020}.
Figure~\ref{fig:strecoil401} shows the 
distribution of recoil mass against $D_s^-\to K^-K^+\pi^-$ tags in the $4180$ dataset.  All
$e^+e^-\to D^{*\pm}_sD^\mp_s$ events accumulate near $m_{D_s^*} = 2112.2$~MeV$/c^2$~\cite{pdg2020}, 
with the direct component populating the central peak and the indirect component distributed more broadly.
\noindent 
We require  $2050<M_{\text{rec}}<2195$~MeV$/c^2$ for the $4180$ data
to ensure that events are consistent with $e^+e^-\to D^{*\pm}_sD^\mp_s$.
For other data samples taken at higher $E_{\text{cm}}$, the tails of the indirect
$D_s$ events extend more widely. We expand the selected $M_{\text{rec}}$ range to maintain roughly
constant $KK\pi$ tag efficiencies for different  values of $E_{\text{cm}}$,
except for the $4230$ data, whose energy
is above the threshold for production of
$D_s^*D_s^*$.
In this case we require $2040<M_{\text{rec}}<2220$~MeV$/c^2$
to suppress $D_s$ from $D_s^*D_s^*$ events, which mostly have $M_{\text{rec}}>2230$~MeV$/c^2$.

When multiple reconstructed candidates are found for a given $D_s$ tag mode and electric charge,
we keep only the one with the best $D^*_s\to\gamma D_s$ photon candidate.
In the $D_s^*$ rest frame, the emitted photon energy is monochromatic, with energy 
$(m^2_{D_s^*}c^4 - m^2_{D_s}c^4)/(2m_{D_s^*}c^2)=138.9$~MeV.
Once the tag-side $D_s^-$ is reconstructed, we loop over the remaining photon candidates
to construct the four-momentum of the $D_s^*$ candidate based on two hypotheses.
For the first hypothesis, we assume that the tag is direct, the photon is on the signal side, and
$p(D^{*+}_s) = p_{e^+e^-} - p_{\rm{tag}}$.
For the second, we assume that the tag is indirect, the photon is on the tag side, and
$p(D^{*-}_s) = p_\gamma + p_{\rm{tag}}$.
We then choose between these the combination
that gives $E_\gamma$ closest to $138.9$~MeV in the $D_s^{*}$ rest frame and use this
in the rest of our analysis.
Additionally, we require $119<E_\gamma<149$~MeV to suppress
backgrounds further.  This range is selected
 based on MC studies of the signal side, as is
described in Sec.~\ref{sec:dtana}. The resulting optimized photon selection efficiency is $\sim90\%$.
While we do not use information about the transition photon
in determining the tag-side yields, we perform the photon reconstruction and apply the additional
selection on $E_\gamma$ at this point to minimize the systematic uncertainty due to the
best-tag selection.  The effect on the signal efficiency of selecting the transition photon
is minimal due to the simple decay topology of the signal side, with only one charged track.  

To determine
the tag yields, we perform an unbinned maximum likelihood fit
to $M_{\text{inv}}(D^-_s)$ in the range
$1900<M_{\text{inv}}(D^-_s)<2030$~MeV$/c^2$. The signal functions are based on 
distributions from MC simulations,
obtained by the Gaussian kernel estimation method~\cite{keypdf},
and convolved with a Gaussian function to account for differing resolution 
between data and MC samples.  In practice it is difficult to float the
width of the Gaussian function
in fits for 
tag modes with larger backgrounds in smaller samples ({e.g., $4220$ data).
Because of this, we assume that the relative difference between data and MC samples is consistent 
among the different datasets and simultaneously fit to all six samples, 
sharing a single convolved Gaussian function for a given tag mode.  We estimate possible systematic
uncertainty associated with this assumption in Sec.~\ref{sec:systfit}.

Backgrounds in the invariant-mass fits are represented by low-degree Chebyshev polynomials
(first to third, depending on the tag mode).  Figure~\ref{fig:STFIT_DATA} shows fits to the 
$M_{\text{inv}}(D^-_s)$ distributions of the thirteen tag modes for the $4180$ dataset.
In the fits of tag modes $D^-_s\to K^0_SK^-$ and $D^-_s\to K^-K^+\pi^-$ MC simulations predict 
small peaking backgrounds from $D^-\to K^0_S\pi^-$ and $D^-\to K^+\pi^-\pi^-$, 
which are taken into account in the fits.  Table~\ref{tab:xyzSTeff} shows the
$ST$ efficiencies, $\epsilon_{ST}$. The corresponding
$ST$ yields from data are
also shown in Table~\ref{tab:xyzstyields}.

\begin{figure*}[htbp]
  \centering
  \includegraphics[keepaspectratio=true,width=6.90in,angle=0]{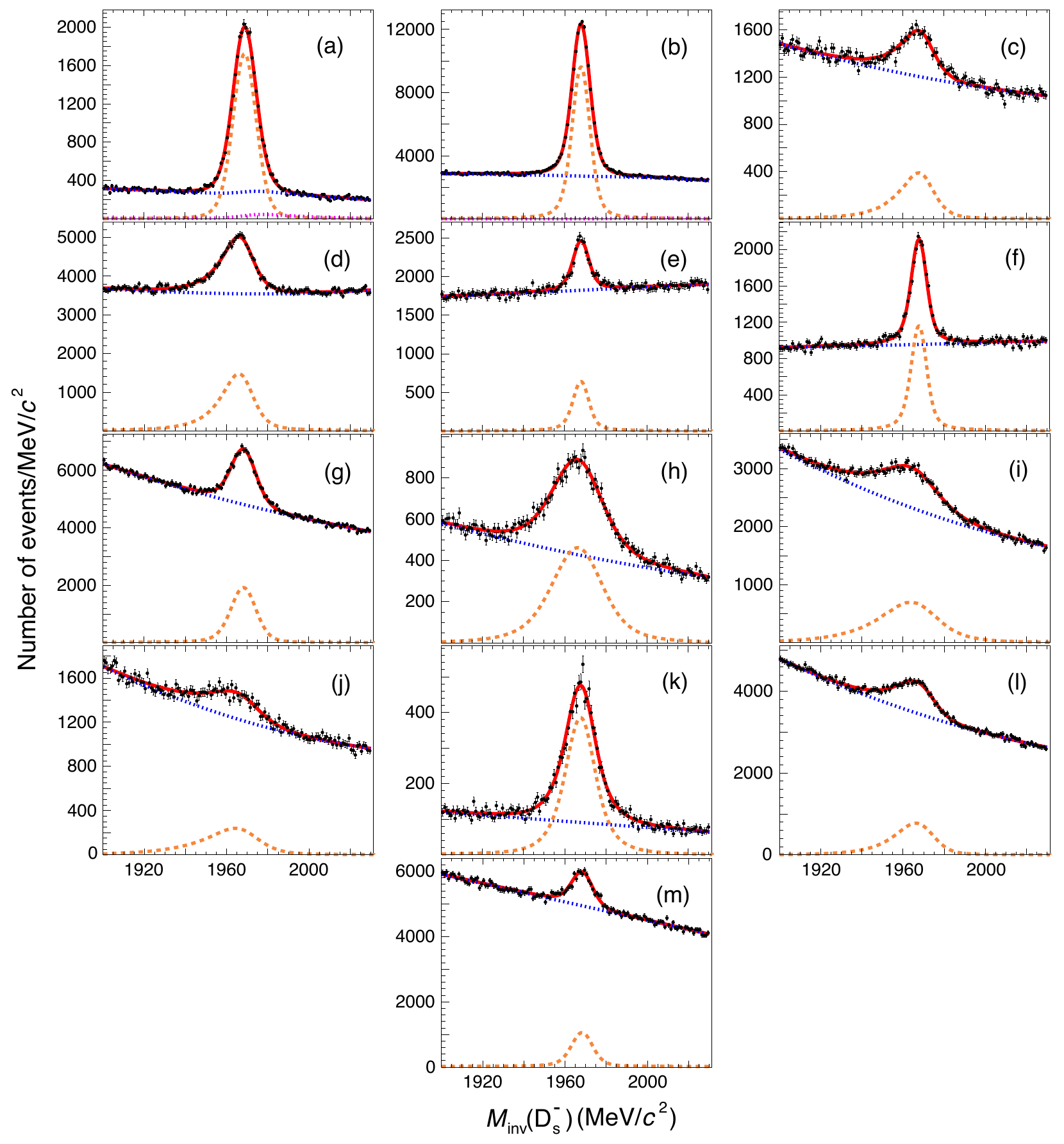} \\
  \caption{Fits to $D_s^-$ invariant-mass distributions in the $4180$ data for
    the following thirteen $ST$ decay modes:
    (a) $K^0_SK^-$, (b) $K^-K^+\pi^-$, (c) $K^0_SK^-\pi^0$, (d) $K^-K^+\pi^-\pi^0$,
    (e) $K^0_SK^-\pi^+\pi^-$, (f) $K^0_SK^+\pi^-\pi^-$, (g) $\pi^-\pi^+\pi^-$, (h) $\pi^-\eta$,
    (i) $\rho^-\eta$, (j) $\rho^-\eta_{3\pi}$, (k) $\pi^-\eta'_{\pi\pi\eta}$, (l) $\pi^-\eta'_{\gamma\rho}$, and
    (m) $K^-\pi^+\pi^-$. Points are data, while red-solid lines represent the total fits,
    blue-dotted lines are the fitted background shapes, and the orange-dashed lines correspond to the fitted signal shapes.
    The magenta-dotted lines only seen in the two tag modes,  (a) $K^0_SK^-$ and (b) $K^-K^+\pi^-$, are the fitted
  nonsmooth background shapes representing $D^-\to K^0_S\pi^-$ and $D^-\to K^+\pi^-\pi^-$, respectively.  }
\label{fig:STFIT_DATA}
\end{figure*}

\begin{table*}[htb]
  \caption{$ST$ reconstruction efficiencies ($\epsilon_{ST}$ in $\%$) with their statistical uncertainties for the 
thirteen tag modes and six data samples.  Efficiencies do not include the following intermediate-state branching fractions: 
$K^0_S\to\pi^+\pi^-$,
$\pi^0\to\gamma\gamma$,
$\eta\to\gamma\gamma$,
$\eta_{3\pi}\to\pi^+\pi^-\pi^0$,
$\eta'_{\pi\pi\eta}\to\pi^+\pi^-\eta$,
$\eta'_{\gamma\rho}\to\gamma\rho^0$, and $\rho\to\pi\pi$.
}
\label{tab:xyzSTeff}
\begin{center}
\scalebox{1.00}
{
  \begin{tabular}{c  c  c  c  c  c  c}
    \hline\hline
 Tag mode & $4180$ & $4190$ & $4200$ & $4210$ &  $4220$ &  $4230$ \\
\hline
 {\ensurestackMath{
   \alignCenterstack{
  K^0_SK^-\cr 
  K^-K^+\pi^-\cr 
  K^0_SK^-\pi^0\cr
  K^-K^+\pi^-\pi^0\cr
  K^0_SK^-\pi^+\pi^-\cr
  K^0_SK^+\pi^-\pi^-\cr
  \pi^-\pi^+\pi^-\cr
  \pi^-\eta_{\gamma\gamma}\cr
  \rho^-\eta_{\gamma\gamma}\cr
  \rho^-\eta_{3\pi}\cr
  \pi^-\eta'_{\pi\pi\eta}\cr
  \pi^-\eta'_{\gamma\rho}\cr
  K^-\pi^+\pi^-}}} & 
   {\ensurestackMath{
   \alignCenterstack{
 38.5\pm&0.1\cr 
 34.3\pm&0.1\cr 
 14.3\pm&0.2\cr 
  9.2\pm&0.1\cr 
 17.1\pm&0.2\cr 
 19.0\pm&0.1\cr 
 42.6\pm&0.2\cr 
 35.5\pm&0.2\cr 
 13.6\pm&0.1\cr 
 6.1\pm&0.1\cr 
 20.3\pm&0.1\cr 
 27.3\pm&0.3\cr 
 35.9\pm&0.3}}} &
  {\ensurestackMath{
   \alignCenterstack{
38.1\pm&0.3\cr 
34.1\pm&0.1\cr 
14.6\pm&0.5\cr 
9.1\pm&0.1\cr 
16.9\pm&0.5\cr 
18.8\pm&0.2\cr 
41.7\pm&0.5\cr 
35.8\pm&0.5\cr 
13.9\pm&0.3\cr 
5.6\pm&0.4\cr 
19.8\pm&0.2\cr 
27.5\pm&0.7\cr 
35.0\pm&0.8}}} &
  {\ensurestackMath{
   \alignCenterstack{
38.6\pm&0.3\cr
34.4\pm&0.1\cr
14.3\pm&0.5\cr
 9.2\pm&0.1\cr
16.7\pm&0.5\cr
19.2\pm&0.2\cr
42.3\pm&0.6\cr
35.5\pm&0.6\cr
13.6\pm&0.3\cr
5.6\pm&0.4\cr
20.1\pm&0.2\cr
26.6\pm&0.8\cr
34.2\pm&0.8}}} &
  {\ensurestackMath{
   \alignCenterstack{
38.7\pm&0.2\cr
34.1\pm&0.2\cr
13.8\pm&0.5\cr
 9.5\pm&0.1\cr
16.0\pm&0.5\cr
19.1\pm&0.2\cr
41.8\pm&0.5\cr
33.9\pm&0.6\cr
13.4\pm&0.3\cr
6.3\pm&0.4\cr
20.0\pm&0.2\cr
26.4\pm&0.7\cr
34.0\pm&0.8}}} &
  {\ensurestackMath{
   \alignCenterstack{
38.3\pm&0.3\cr
34.4\pm&0.1\cr
13.8\pm&0.6\cr
 9.1\pm&0.2\cr
16.6\pm&0.6\cr
18.9\pm&0.3\cr
41.8\pm&0.7\cr
33.8\pm&0.8\cr
13.5\pm&0.4\cr
5.6\pm&0.5\cr
20.1\pm&0.3\cr
28.3\pm&1.0\cr
33.3\pm&1.0}}} &
  {\ensurestackMath{
   \alignCenterstack{
39.0\pm&0.3\cr
34.8\pm&0.1\cr
14.7\pm&0.5\cr
 9.5\pm&0.1\cr
17.8\pm&0.5\cr
19.8\pm&0.2\cr
43.1\pm&0.6\cr
35.7\pm&0.6\cr
14.5\pm&0.4\cr
5.9\pm&0.4\cr
20.2\pm&0.2\cr
28.0\pm&0.8\cr
         34.7\pm&0.8}}} \\
    \hline\hline
\end{tabular}
}
\end{center}
\end{table*}

\begin{table*}[htb]
  \caption{Measured $ST$ yields ($N_{ST}$) for each tag mode and their sums over tag modes 
(``SUM'') for each data sample, in units of $10^3$.  The uncertainties shown are only statistical.
}
\label{tab:xyzstyields}
\begin{center}
\scalebox{1.0}
{
  \begin{tabular}{c  c  c  c  c  c  c }
    \hline\hline
 Tag mode & $4180$ & $4190$ & $4200$ & $4210$ &  $4220$ &  $4230$ \\
\hline
 {\ensurestackMath{
   \alignCenterstack{
  K^0_SK^-\cr 
  K^-K^+\pi^-\cr 
  K^0_SK^-\pi^0\cr
  K^-K^+\pi^-\pi^0\cr
  K^0_SK^-\pi^+\pi^-\cr
  K^0_SK^+\pi^-\pi^-\cr
  \pi^-\pi^+\pi^-\cr
  \pi^-\eta_{\gamma\gamma}\cr
  \rho^-\eta_{\gamma\gamma}\cr
  \rho^-\eta_{3\pi}\cr
  \pi^-\eta'_{\pi\pi\eta}\cr
  \pi^-\eta'_{\gamma\rho}\cr
  K^-\pi^+\pi^-}}} & 
   {\ensurestackMath{
   \alignCenterstack{
26.2\pm&0.2\cr 
120.6\pm&0.6\cr 
 9.7\pm&0.4\cr 
35.1\pm&0.8\cr 
 7.8\pm&0.3\cr 
14.0\pm&0.2\cr 
33.1\pm&0.6\cr 
16.0\pm&0.5\cr 
28.5\pm&0.9\cr 
 8.6\pm&0.6\cr 
 8.4\pm&0.1\cr 
20.8\pm&0.8\cr 
16.0\pm&0.5}}} &
  {\ensurestackMath{
   \alignCenterstack{
4.1\pm&0.1\cr 
19.0\pm&0.2\cr 
1.5\pm&0.2\cr 
6.3\pm&0.3\cr 
1.2\pm&0.1\cr 
2.2\pm&0.1\cr 
5.4\pm&0.2\cr 
2.4\pm&0.2\cr 
3.9\pm&0.4\cr 
1.4\pm&0.2\cr 
1.4\pm&0.1\cr 
3.3\pm&0.3\cr 
2.6\pm&0.2}}} &
  {\ensurestackMath{
   \alignCenterstack{
4.1\pm&0.1\cr
18.6\pm&0.2\cr
1.7\pm&0.2\cr
5.6\pm&0.3\cr
1.3\pm&0.1\cr
2.2\pm&0.1\cr
5.2\pm&0.2\cr
2.4\pm&0.2\cr
4.5\pm&0.4\cr
1.2\pm&0.2\cr
1.3\pm&0.1\cr
3.2\pm&0.3\cr
2.3\pm&0.2}}} &
  {\ensurestackMath{
   \alignCenterstack{
3.6\pm&0.1\cr
17.3\pm&0.2\cr
1.3\pm&0.2\cr
5.2\pm&0.3\cr
1.0\pm&0.1\cr
1.9\pm&0.1\cr
4.3\pm&0.2\cr
2.4\pm&0.2\cr
4.3\pm&0.4\cr
1.1\pm&0.2\cr
1.2\pm&0.1\cr
3.0\pm&0.3\cr
2.3\pm&0.2}}} &
  {\ensurestackMath{
   \alignCenterstack{
3.0\pm&0.1\cr
15.1\pm&0.2\cr
0.9\pm&0.2\cr
4.9\pm&0.3\cr
0.9\pm&0.1\cr
1.9\pm&0.1\cr
3.8\pm&0.2\cr
2.0\pm&0.2\cr
3.5\pm&0.4\cr
0.7\pm&0.2\cr
1.0\pm&0.1\cr
2.5\pm&0.3\cr
1.7\pm&0.2}}} &
  {\ensurestackMath{
   \alignCenterstack{
5.5\pm&0.1\cr
25.2\pm&0.2\cr
2.1\pm&0.2\cr
7.5\pm&0.3\cr
1.5\pm&0.1\cr
2.8\pm&0.1\cr
6.7\pm&0.3\cr
3.2\pm&0.2\cr
6.1\pm&0.5\cr
2.1\pm&0.4\cr
1.7\pm&0.1\cr
4.7\pm&0.4\cr
3.4\pm&0.2}}} \\
\hline
SUM & {\ensurestackMath{
   \alignCenterstack{
      344.8\pm2.0}}} &
  {\ensurestackMath{
   \alignCenterstack{
54.7\pm0.7}}} &
  {\ensurestackMath{
   \alignCenterstack{
53.8\pm0.8}}} &
  {\ensurestackMath{
   \alignCenterstack{
48.8\pm0.8}}} &
  {\ensurestackMath{
   \alignCenterstack{
42.1\pm0.7}}} &
  {\ensurestackMath{
   \alignCenterstack{
                72.5\pm1.0}}} \\
    \hline\hline
\end{tabular}
}
\end{center}
\end{table*}

\subsection{\boldmath Selection of $D_s^+\to\ell^+\nu_\ell$ events}\label{sec:dtana}

True $D_s^+ \rightarrow \ell^+\nu_\ell$ signal events include either $D_s^+ \to \mu^+\nu_\mu$ or
$D_s^+ \to \tau^+ (\to \pi^+\bar{\nu}_\tau) \nu_\tau$ accompanying a $D_s^-$ hadronic tag.
Signal selection begins with the requirement that there be only one additional track that is unused in the reconstruction of the tag ($N_{tk}=1$).
The corresponding particle must have electric charge opposite to the tag and satisfy the pion PID criteria described in Sec. ~\ref{sec:STselect}.
(The pion PID response closely approximates that for a muon because pions and muons are charged particles with similar masses.)
It is significant that the systematic uncertainty arising from the PID requirement in the signal selection 
does not cancel, as it does for tag reconstruction.  We study control samples of
$D_s^-\to K^-K^+\pi^-$, $D^0\to K^-\pi^+$ and $D^0\to K^-\pi^-\pi^+\pi^+$ for all datasets and observe 
differences in PID efficiencies between data and MC samples of about $1\%$.
Here the $D^0$ sample is obtained via $e^+e^-\to D^{*+}(\to\pi^+D^0)D^{(*)-}$.
Corrections  are applied to MC-determined 
efficiencies.

We split the signal-track sample into two parts based on the energy-deposit properties of muons and
pions in the EMC, as was done previously in similar analyses~\cite{CLEODs,BES3Ds}.
Candidates with signal-track energy deposit satisfying $E_{\rm{EMC}} \le 300$~MeV
are classified as $\mu$-like and the remainder as $\pi$-like.  Based on MC simulation, we estimate that 
the $\mu$-like sample includes $\sim 99\%$ of $D^+_s\to\mu^+\nu_\mu$ events and 
$\sim58\%$ of $D^+_s\to\tau^+(\to \pi^+\bar{\nu}_\tau)\nu_\tau$ events.

As mentioned in Sec.~\ref{sec:STselect}, we impose the additional requirement on signal candidates
from $D_s^*\to\gamma D_s$ that the selected photon have an energy in the $D_s^*$ rest frame that satisfies $119<E_\gamma<149$~MeV.  This criterion is optimized based on a detailed MC study. Distributions of $E_\gamma$
  for MC and data are shown in Fig.~\ref{fig:cutsegamma}.
Here the input $\b(D_s^+\to\mu^+\nu_\mu)$ and $\b(D_s^+\to\tau^+\nu_\tau)$ in our MC
samples are $5.38\times10^{-3}$ and $5.54\times10^{-2}$, respectively.

\begin{figure}[htbp]
  \centering
  \includegraphics[keepaspectratio=true,width=3.1in,angle=0]{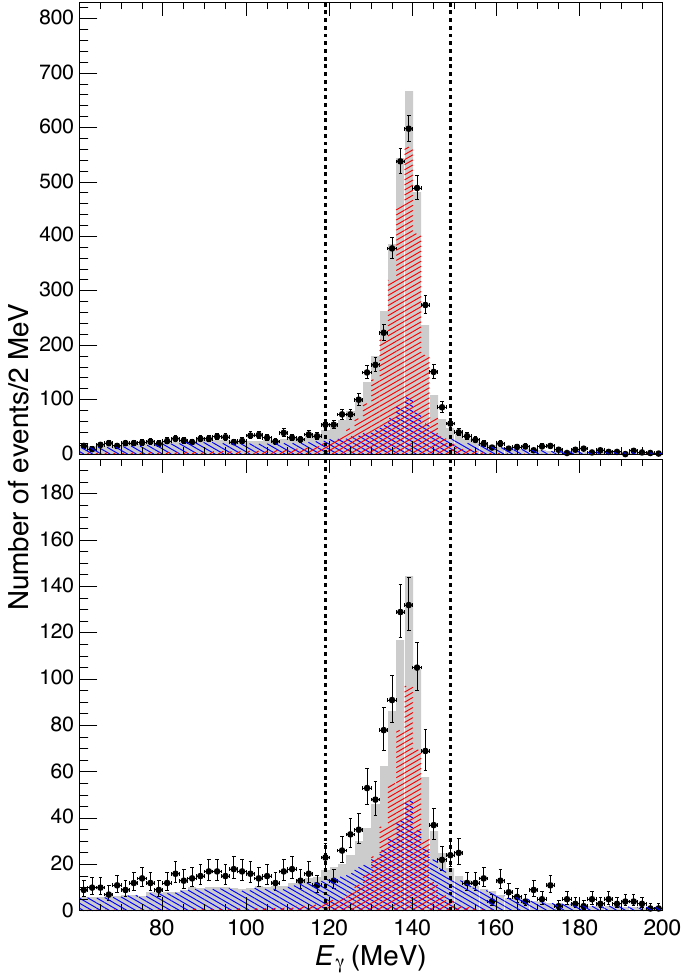}
\caption{$E_\gamma$ spectra in the $D^*_s$ rest frame for $\mu$-like (top)
and $\pi$-like (bottom) samples summed over the six $E_{\text{cm}}$ values (weighted by integrated 
luminosity).  The gray-shaded histograms represent the scaled $40\times$ MC predictions, while the
crosshatched
histograms show the signals (red, bottom-left to top-right)
and the backgrounds (blue, top-left to bottom-right).
The black points with error bars 
are the corresponding measurements from data, and the dashed vertical lines show our nominal selection
requirement of $119<E_\gamma<149$~MeV.
}
\label{fig:cutsegamma}
\end{figure}

We suppress candidate signal events that are not true $D_s^+\to\ell^+\nu_\ell$ by considering three additional
variables that are sensitive to unreconstructed particles.  The first is $\cos{\theta_\text{miss}}$, 
where $\theta_\text{miss}$ 
is the polar angle of $\vec{p}_{\rm{miss}} = -\vec{p}_{\rm{tag}} - \vec{p}_\gamma - \vec{p}_{\mu/\pi}$ in the 
$e^+e^-$ center-of-mass frame.  We require $|\cos{\theta_\text{miss}}|<0.9$ to ensure that $\vec{p}_{\rm{miss}}$ 
points into the fiducial volume of our detector.  The second variable is $ANG$, the opening angle between 
$\vec{p}_{\rm{miss}}$ and the most energetic unused EMC shower.  Events with a $K^0_L$ or
an energetically asymmetric decay from $\pi^0\to\gamma\gamma$ or $\eta\to\gamma\gamma$
tend to leave detectable EMC energy deposits near the $\vec{p}_{\rm{miss}}$
direction.  We require $ANG>40\degree$ to suppress such events.  The third variable is $E^\text{max}_\text{neu}$, 
the maximum unused EMC shower energy. Requiring $E^\text{max}_\text{neu} < 300$~MeV helps to ensure that
nothing energetic is unaccounted for in selected events.

\subsection{\boldmath Fit}\label{sec:dtfit}

\subsubsection{Fitting to simulated samples}

We infer the presence of
neutrinos in the final states from the event missing mass-squared,
$M^2_{\rm{miss}}$ $= E^2_{\rm{miss}} - |\vec{p}_{\rm{miss}}c|^2$, where
$E_{\rm{miss}} = E_{\rm{cm}} -\sqrt{|\vec{p}_{\rm{tag}}c|^2+m^2_{D_s}c^4} - E_\gamma -E_{\mu/\pi}$ 
is computed in the $e^+e^-$ center-of-mass frame, and $m_{D_s}$ is fixed to the known 
value~\cite{pdg2020}.  We determine
the $\b(D_s^+\to\ell^+\nu)$ signal yields with a simultaneous unbinned 
maximum likelihood fit to the two-dimensional distributions of the tag-side invariant mass 
$M_{\text{inv}}(D^-_s)$ versus $M^2_{\rm{miss}}$ for the six $E_{\text{cm}}$ samples.  The fit range on the 
$M_{\text{inv}}(D^-_s)$ axis is the same as we use to fit to the tag-side $D^-_s$, 
$1900<M_{\text{inv}}(D^-_s)<2030$~MeV$/c^2$, while we choose 
$-0.2<M^2_{\rm{miss}}<0.2$~$({\rm GeV}/c^2)^2$ on the other axis to avoid peaking backgrounds
in $M^2_{\rm{miss}}$$>0.2$~$({\rm GeV}/c^2)^2$ coming from $D_s^+\to K^0\pi^+$ when the $K^0$ is undetected.

Background processes contributing $D_s^+\to\ell^+\nu_\ell$ candidate signal events can be classified
in five major categories.  (1) The tag-side $D^-_s$ is misreconstructed, but everything else
is correct, producing smooth distributions in $M_{\text{inv}}$, but peaks in $M^2_{\rm{miss}}$.
(2) The photon from $D^*_s\to\gamma D_s$ is misreconstructed, but everything else is correct, leading to
  peaking sharply in $M_{\text{inv}}$ and broadly in $M^2_{\rm{miss}}$.
(3) The signal-side track is not from $D^+_s\to (\mu^+/\tau^+)\nu$, but everything else
is correct, giving peaks in $M_{\text{inv}}$ and smooth distributions in $M^2_{\rm{miss}}$.
(4) The reconstructed track is from $D^+_s\to \tau^+\nu_\tau$, but not from
$\tau^+\to\pi^+\bar{\nu}_\tau$,
and everything else is correct, with peaks in $M_{\text{inv}}$ but smooth distributions in $M^2_{\rm{miss}}$.  (Note that, while such events
are not analyzed correctly, they are not background and we float this contribution with the  
$\tau^+\to\pi^+\bar{\nu}_\tau$ signal.)  (5) All remaining background, both charm 
and noncharm, which has a smooth continuous shape.  Two-dimensional background 
distributions for fitting are constructed based on products of one-dimensional probability
distribution functions (PDFs) for $M_{\text{inv}}(D^-_s)$ and $M^2_{\rm{miss}}$.

Fits are performed with MC-based shapes for signals,
similarly extracted by the kernel estimation method~\cite{keypdf},
and the following distributions for the five 
background components.
(1) Wrong tag: first-order Chebyshev polynominal for $M_{\text{inv}}(D^-_s)$ and an MC-based 
shape for $M^2_{\rm{miss}}$.
(2) Wrong photon: MC-based shapes for both $M_{\text{inv}}(D^-_s)$ and $M^2_{\rm{miss}}$.
(3) Wrong track: MC-based shapes for both $M_{\text{inv}}(D^-_s)$ and $M^2_{\rm{miss}}$.
(4) $\tau$ decays to final states other than $\pi^+\bar{\nu}_\tau$: MC-based shapes for both 
$M_{\text{inv}}(D^-_s)$ and $M^2_{\rm{miss}}$.
(5) Remainder: first-order Chebyshev polynominal for $M_{\text{inv}}(D^-_s)$ and first-order 
exponential polynomial ($\propto e^{C\cdot \text{$M^2_{\rm{miss}}$}}$) for $M^2_{\rm{miss}}$.
To compensate for differences between the simulated detector response and data,
each PDF [$M_{\text{inv}}(D^-_s)$ and $M^2_{\rm{miss}}$] is convolved with a Gaussian function that is shared over the
six data samples, assuming that the data/MC differences do not depend on $E_{\text{cm}}$ and 
running conditions.

Normalizations of the five background components are fixed whenever an absolute estimate is
possible and otherwise are allowed to float freely in fitting, as follows.
(1) Wrong tag: fixed based on the MC-estimated ratio of the correctly reconstructed yields to the 
wrongly reconstructed yields.  (We float this component in assessing associated systematic uncertainties.)
(2) Wrong photon: ratio to the signal component is fixed according to the MC estimation.
(3) Wrong track: fixed according to a MC estimation for which the MC sample is scaled to the $N_{ST}$ 
observed in data.
(4) $\tau$ decays to final states other than $\pi^+\bar{\nu}_\tau$: constrained to the signal
yield, $D^+_s\to\tau^+(\to\pi^+\bar{\nu}_\tau)\nu_\tau$.
(5) Remainder: floated freely.

Normalizations of the two signal components ($D_s^+\to\mu\nu_\mu$ and $\tau^+\nu_\tau$) are floated freely,
except for the constraints introduced by the simultaneous fit and the ratio of the yields between
the $\mu$-like and $\pi$-like samples. This ratio is fixed for each of the six datasets
based on MC estimations.

The means of the convolved Gaussian functions in fitting $M_{\text{inv}}(D^-_s)$ and $M^2_{\rm{miss}}$ are floated for both data and MC simulations.
The widths are floated in fitting to data and are fixed to a negligibly small value in fitting to MC samples.
Because the $M_{\text{inv}}(D^-_s)$ resolution depends on the $D^-_s$ decay mode, 
the $M_{\text{inv}}(D^-_s)$ PDF is obtained as a sum of PDFs for all tag modes, weighted by the 
observed $N_{ST}$.  We obtain the $M^2_{\rm{miss}}$ PDFs in a similar way,
as MC simulation predicts weak tag-mode dependence.
The coefficients of the Chebyshev polynomials for backgrounds (1) 
and (5) are shared because MC simulations demonstrate that the statistical sensitivity of our data is insufficient 
to distinguish their slopes.

We determine $\epsilon_{DT}^{\ell\nu}$ for each of the thirteen tag modes by counting the
reconstructed candidates that match the MC-generated true signal particles.
The resultant efficiencies are shown in
Tables~\ref{tab:xyzDTeffmu} and \ref{tab:xyzDTeffta}.

\begin{table*}[bth]
  \caption{$DT$ reconstruction efficiencies (in $\%$) with statistical uncertainties for $D_s^+\to\mu^+\nu_\mu$ signal 
events in each of the thirteen tag modes.  These efficiencies do not include the following intermediate-state branching fractions: 
$K^0_S\to\pi^+\pi^-$,
$\pi^0\to\gamma\gamma$,
$\eta\to\gamma\gamma$,
$\eta_{3\pi}\to\pi^+\pi^-\pi^0$,
$\eta'_{\pi\pi\eta}\to\pi^+\pi^-\eta$,
$\eta'_{\gamma\rho}\to\gamma\rho^0$, and $\rho\to\pi\pi$.
}
\label{tab:xyzDTeffmu}
\begin{center}
\scalebox{1.00}
{
  \begin{tabular}{c  c  c  c  c  c  c}
    \hline\hline
 Tag mode & $4180$ & $4190$ & $4200$ & $4210$ &  $4220$ &  $4230$ \\
\hline
 {\ensurestackMath{
   \alignCenterstack{
  K^0_SK^-\cr 
  K^-K^+\pi^-\cr 
  K^0_SK^-\pi^0\cr
  K^-K^+\pi^-\pi^0\cr
  K^0_SK^-\pi^+\pi^-\cr
  K^0_SK^+\pi^-\pi^-\cr
  \pi^-\pi^+\pi^-\cr
  \pi^-\eta_{\gamma\gamma}\cr
  \rho^-\eta_{\gamma\gamma}\cr
  \rho^-\eta_{3\pi}\cr
  \pi^-\eta'_{\pi\pi\eta}\cr
  \pi^-\eta'_{\gamma\rho}\cr
  K^-\pi^+\pi^-}}} & 
   {\ensurestackMath{
   \alignCenterstack{
 23.23\pm&0.12\cr 
 19.84\pm&0.05\cr 
 10.54\pm&0.09\cr 
  6.88\pm&0.03\cr 
 11.14\pm&0.11\cr 
 12.13\pm&0.09\cr 
 26.88\pm&0.12\cr 
 25.24\pm&0.15\cr 
 12.54\pm&0.05\cr 
 5.35\pm&0.05\cr 
 13.48\pm&0.12\cr 
 19.43\pm&0.11\cr 
 23.31\pm&0.16}}} &
  {\ensurestackMath{
   \alignCenterstack{
23.06\pm&0.17\cr 
19.63\pm&0.07\cr 
10.29\pm&0.13\cr 
6.78\pm&0.04\cr 
10.82\pm&0.16\cr 
11.88\pm&0.13\cr 
26.64\pm&0.17\cr 
24.80\pm&0.22\cr 
12.40\pm&0.07\cr 
5.20\pm&0.07\cr 
13.35\pm&0.18\cr 
18.91\pm&0.16\cr 
22.72\pm&0.23}}} &
  {\ensurestackMath{
   \alignCenterstack{
23.32\pm&0.18\cr
19.63\pm&0.07\cr
10.46\pm&0.13\cr
7.01\pm&0.04\cr
11.24\pm&0.17\cr
12.02\pm&0.13\cr
26.56\pm&0.17\cr
24.87\pm&0.22\cr
12.32\pm&0.07\cr
5.29\pm&0.07\cr
13.19\pm&0.18\cr
19.02\pm&0.16\cr
23.41\pm&0.23}}} &
  {\ensurestackMath{
   \alignCenterstack{
22.33\pm&0.17\cr
19.25\pm&0.07\cr
10.43\pm&0.13\cr
6.89\pm&0.04\cr
10.61\pm&0.16\cr
11.78\pm&0.13\cr
26.07\pm&0.17\cr
24.53\pm&0.22\cr
12.29\pm&0.07\cr
5.12\pm&0.07\cr
13.17\pm&0.18\cr
19.05\pm&0.16\cr
22.57\pm&0.23}}} &
  {\ensurestackMath{
   \alignCenterstack{
22.64\pm&0.18\cr
19.52\pm&0.07\cr
10.11\pm&0.13\cr
6.92\pm&0.05\cr
10.94\pm&0.17\cr
11.81\pm&0.13\cr
26.12\pm&0.17\cr
24.48\pm&0.22\cr
12.25\pm&0.07\cr
5.05\pm&0.07\cr
13.33\pm&0.18\cr
18.84\pm&0.16\cr
22.51\pm&0.23}}} &
  {\ensurestackMath{
   \alignCenterstack{
23.80\pm&0.18\cr
20.37\pm&0.07\cr
10.80\pm&0.13\cr
7.36\pm&0.05\cr
11.70\pm&0.17\cr
12.72\pm&0.13\cr
27.71\pm&0.18\cr
25.41\pm&0.22\cr
12.80\pm&0.08\cr
5.39\pm&0.07\cr
13.99\pm&0.18\cr
20.04\pm&0.16\cr
          23.82\pm&0.23}}} \\
    \hline\hline
\end{tabular}
}
\end{center}
\end{table*}

\begin{table*}[bth]
  \caption{$DT$ reconstruction efficiencies (in $\%$) with statistical uncertainties for $D_s^+\to\tau^+(\to\pi\bar{\nu}_\tau)\nu_\tau$ 
signal events in each of the thirteen tag modes.
These efficiencies do not include the following intermediate-state branching fractions: $K^0_S\to\pi^+\pi^-$,
$\pi^0\to\gamma\gamma$,
$\eta\to\gamma\gamma$,
$\eta_{3\pi}\to\pi^+\pi^-\pi^0$,
$\eta'_{\pi\pi\eta}\to\pi^+\pi^-\eta$,
$\eta'_{\gamma\rho}\to\gamma\rho^0$, and $\rho\to\pi\pi$.
}
\label{tab:xyzDTeffta}
\begin{center}
\scalebox{1.00}
{
  \begin{tabular}{c  c  c  c  c  c  c}
    \hline\hline
 Tag mode & $4180$ & $4190$ & $4200$ & $4210$ &  $4220$ &  $4230$ \\
\hline
 {\ensurestackMath{
   \alignCenterstack{
  K^0_SK^-\cr 
  K^-K^+\pi^-\cr 
  K^0_SK^-\pi^0\cr
  K^-K^+\pi^-\pi^0\cr
  K^0_SK^-\pi^+\pi^-\cr
  K^0_SK^+\pi^-\pi^-\cr
  \pi^-\pi^+\pi^-\cr
  \pi^-\eta_{\gamma\gamma}\cr
  \rho^-\eta_{\gamma\gamma}\cr
  \rho^-\eta_{3\pi}\cr
  \pi^-\eta'_{\pi\pi\eta}\cr
  \pi^-\eta'_{\gamma\rho}\cr
  K^-\pi^+\pi^-}}} & 
   {\ensurestackMath{
   \alignCenterstack{
 9.79\pm&0.08\cr 
 8.33\pm&0.03\cr 
 4.34\pm&0.06\cr 
  2.83\pm&0.02\cr 
 4.60\pm&0.07\cr 
 5.14\pm&0.06\cr 
 11.26\pm&0.08\cr 
 10.44\pm&0.11\cr 
 5.05\pm&0.03\cr 
 2.11\pm&0.03\cr 
 5.64\pm&0.08\cr 
 8.16\pm&0.07\cr 
 9.71\pm&0.11}}} &
  {\ensurestackMath{
   \alignCenterstack{
9.66\pm&0.12\cr 
8.19\pm&0.05\cr 
4.28\pm&0.08\cr 
2.78\pm&0.03\cr 
4.58\pm&0.11\cr 
4.87\pm&0.09\cr 
11.10\pm&0.12\cr 
10.41\pm&0.16\cr 
5.06\pm&0.05\cr 
2.11\pm&0.04\cr 
5.43\pm&0.12\cr 
8.08\pm&0.11\cr 
9.41\pm&0.16}}} &
  {\ensurestackMath{
   \alignCenterstack{
9.76\pm&0.12\cr
8.30\pm&0.05\cr
4.19\pm&0.08\cr
2.84\pm&0.03\cr
4.54\pm&0.11\cr
4.98\pm&0.09\cr
11.11\pm&0.12\cr
10.19\pm&0.16\cr
4.96\pm&0.05\cr
2.01\pm&0.04\cr
5.45\pm&0.12\cr
8.09\pm&0.11\cr
9.78\pm&0.16}}} &
  {\ensurestackMath{
   \alignCenterstack{
9.32\pm&0.12\cr
7.98\pm&0.05\cr
4.16\pm&0.08\cr
2.79\pm&0.03\cr
4.47\pm&0.11\cr
4.83\pm&0.09\cr
10.77\pm&0.12\cr
10.25\pm&0.16\cr
4.89\pm&0.05\cr
1.95\pm&0.04\cr
5.51\pm&0.12\cr
7.99\pm&0.11\cr
9.77\pm&0.16}}} &
  {\ensurestackMath{
   \alignCenterstack{
9.62\pm&0.12\cr
8.03\pm&0.05\cr
4.09\pm&0.08\cr
2.81\pm&0.03\cr
4.55\pm&0.11\cr
4.89\pm&0.09\cr
10.96\pm&0.12\cr
10.21\pm&0.16\cr
4.91\pm&0.05\cr
2.04\pm&0.04\cr
5.39\pm&0.12\cr
7.94\pm&0.11\cr
9.47\pm&0.16}}} &
  {\ensurestackMath{
   \alignCenterstack{
9.83\pm&0.12\cr
8.46\pm&0.05\cr
4.35\pm&0.08\cr
2.92\pm&0.03\cr
4.80\pm&0.11\cr
5.17\pm&0.09\cr
11.75\pm&0.13\cr
10.14\pm&0.16\cr
5.05\pm&0.05\cr
2.18\pm&0.04\cr
5.66\pm&0.12\cr
8.10\pm&0.11\cr
         9.78\pm&0.16}}} \\
    \hline\hline
\end{tabular}
}
\end{center}
\end{table*}

To validate these $DT$ efficiencies and our overall fitting procedure,
we perform tests on ten independent data-size MC samples
and compare the fitted signal yields and the corresponding branching fractions.
Table~\ref{tab:IOrounds} shows the differences between
the fitted signal yields and the MC-predicted yields
for the data-size samples and for
a $40\times$ sample. We see reasonable
agreement between the fitted and generated yields.

\begin{table}[h]
  \caption{
    Relative difference in $\%$ between the measured signal yields and generated
    numbers for ten ``sets'' of data-sized MC samples, the average of these, and
    a $40\times$ MC sample.
}
\label{tab:IOrounds}
\begin{center}
\scalebox{1.0}
{
  \begin{tabular}{c|| c |  c }
    \hline\hline
        & Relative Difference & Relative Difference \\
  Set & of $\mu\nu$ Yield & of $\tau\nu$ Yield \\
\hline
$01$ & $+0.41\pm2.43$ & $-1.54\pm4.27$ \\
$02$ & $-0.76\pm2.39$ & $+2.09\pm4.07$ \\
$03$ & $+1.09\pm2.43$ & $+1.96\pm4.25$ \\
$04$ & $-1.21\pm2.45$ & $+3.10\pm4.20$ \\
$05$ & $+0.40\pm2.37$ & $-0.87\pm4.37$ \\
$06$ & $-0.09\pm2.38$ & $+3.32\pm4.10$ \\
$07$ & $-2.85\pm2.41$ & $+3.61\pm5.01$ \\
$08$ & $+1.12\pm2.40$ & $+0.25\pm4.21$ \\
$09$ & $+0.29\pm2.34$ & $-1.91\pm4.45$ \\
$10$ & $-1.18\pm2.37$ & $-3.27\pm4.26$ \\
\hline
  Average & $-0.52\pm0.76$ & $+0.57\pm1.36$ \\
  \hline
    $40\times$ & $-0.30\pm0.39$ & $-0.68\pm0.70$\\
    \hline\hline
\end{tabular}
}
\end{center}
\end{table}

Figure~\ref{fig:mcbkgDATA} shows projections of the selected data sample
onto the $M^2_{\rm{miss}}$ and 
$M_{\text{inv}}(D_s^-)$ axes, summed over the six data samples.
Scaled MC distributions are overlaid, with
a breakdown of the components of the MC-simulated background.
There is agreement between data and MC simulation for both signal and background.  Figures~\ref{fig:mcbkgDATAboomom}  and \ref{fig:mcbkgDATAboocos} show comparisons between data and MC samples for signal-track momentum and the cosine of the polar angle, again demonstrating excellent agreement.

  \begin{figure*}[t]
  \centering
   \includegraphics[keepaspectratio=true,width=6.5in,angle=0]{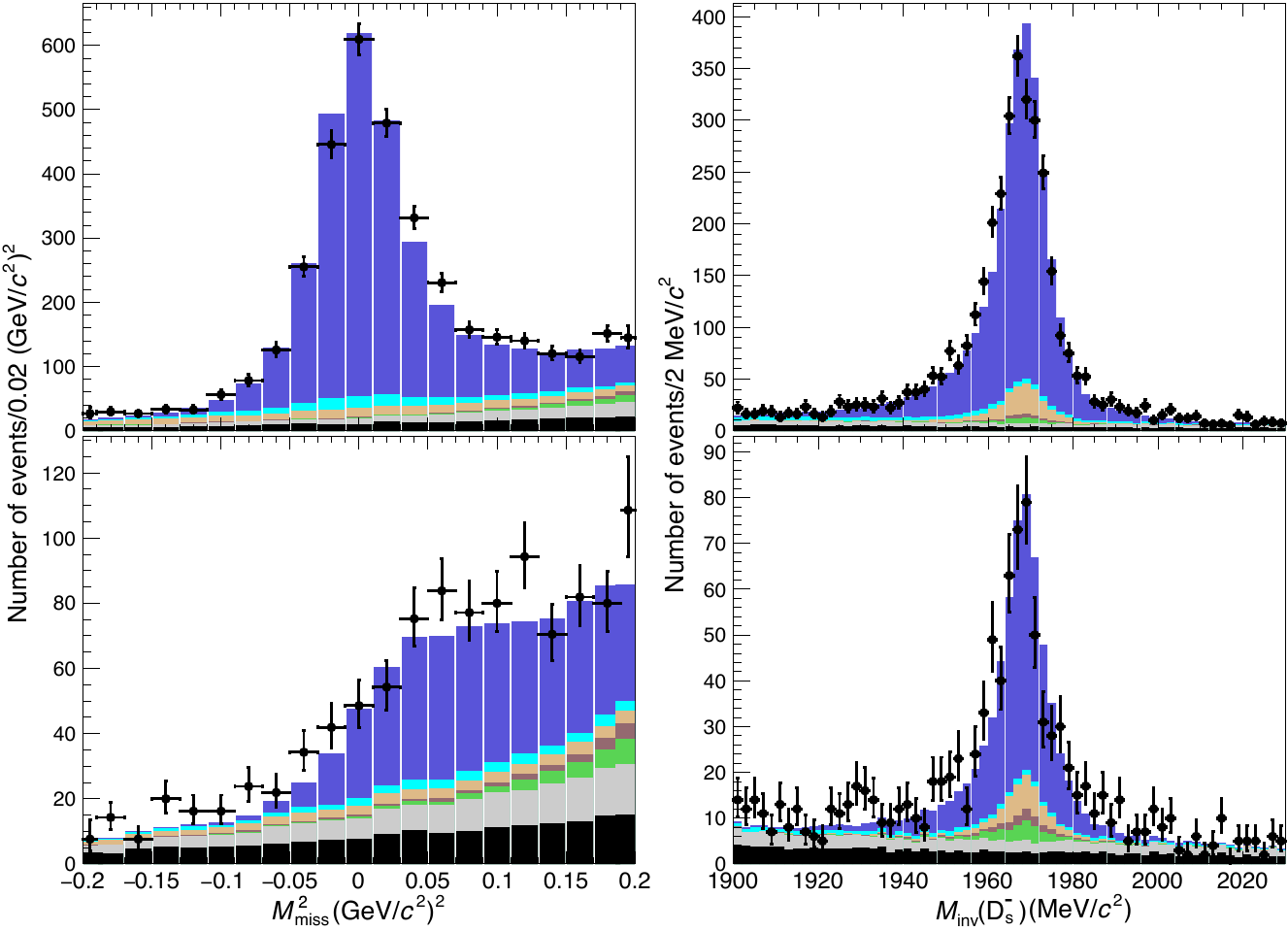}
  \caption{Projections onto the $M^2_{\rm{miss}}$ (left) and $M_{\text{inv}}(D^-_s)$ (right) axes,
  for $\mu$-like (top) and $\pi$-like (bottom) candidates.  Black points are data, 
  summed over all six samples.
  The filled histograms represent the prediction of the $40\times$ MC sample,
  normalized to the integrated luminosity of data, with the
  $D_s^+\to\mu^+\nu_\mu$ and $\tau^+(\to\pi^+\bar{\nu}_\tau)\nu_\tau$ signal on top (blue).
  Background components are stacked below this in the following order from the top to the bottom: 
  tag-side misreconstructed (cyan),
  misreconstructed transition photon (light brown),
  signal side misreconstructed (dark brown),
  $D_s^+\to\tau^+\nu_\tau$, where the $\tau^+$ decays to a final state other than $\pi^+\bar{\nu}_\tau$ (green),
  other charm background (gray), and noncharm sources (black, at bottom).
}
\label{fig:mcbkgDATA}
\end{figure*}

\begin{figure*}[htbp]
  \centering
 \includegraphics[keepaspectratio=true,width=6.5in,angle=0]{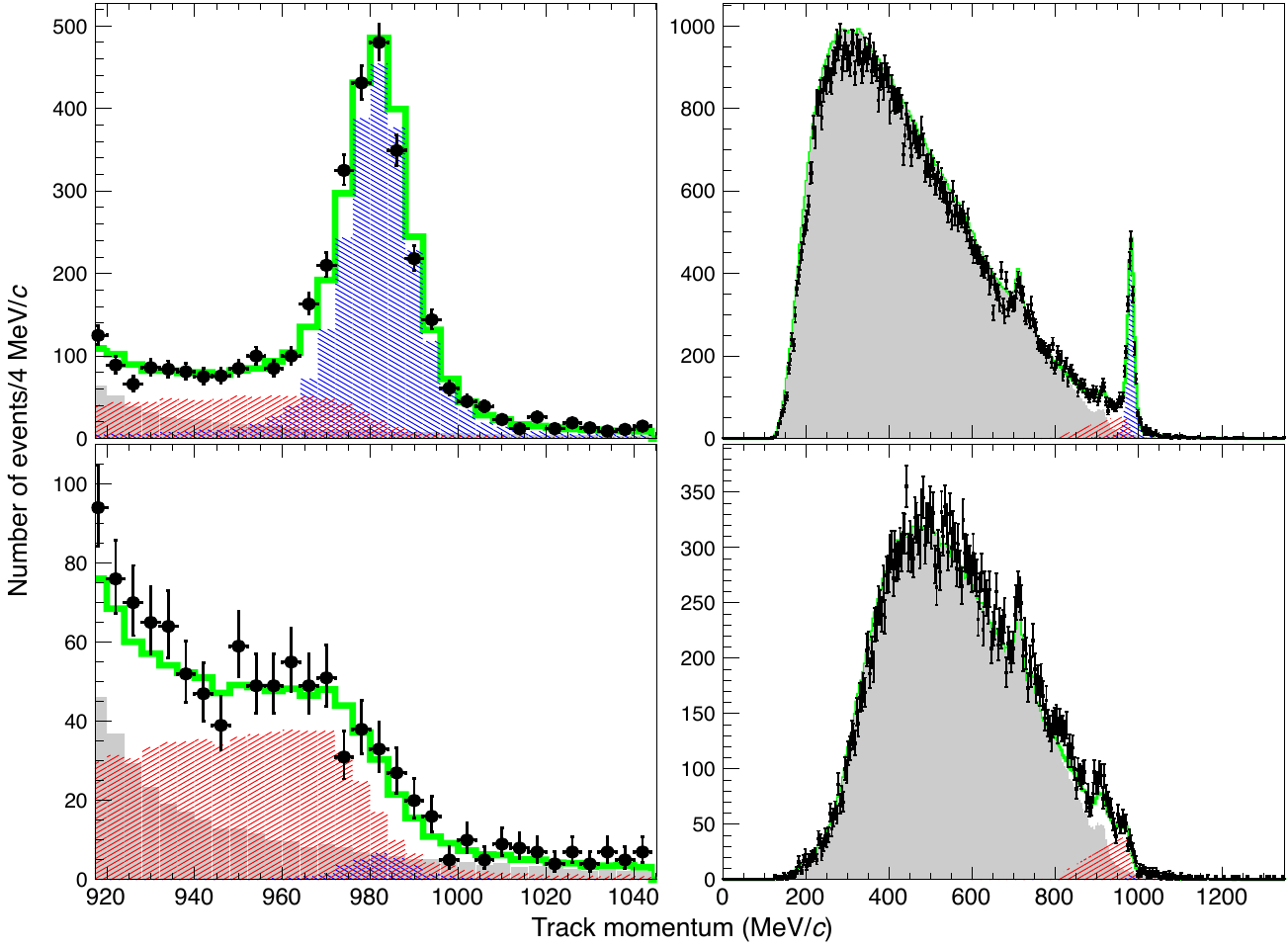}
  \caption{Distributions of the momentum of signal
  candidate tracks in the $D_s^+$ rest frame for the $DT$
  sample with the additional requirement $1900<M_{\text{inv}}(D^-_s)<2030$~MeV$/c^2$
  for the  $\mu$-like (top) and the  $\pi$-like samples (bottom).
  The black points are data and the overlaid histograms represent the $40\times$ MC sample (normalized
  to the integrated luminosity of data), with green for the total,
  gray filled for the total background,
  and crosshatched for the signals,  $D^+_s\to\mu^+\nu_\mu$ (blue, top-left to bottom-right)
  and $D^+_s\to\tau^+(\to\pi^+\bar{\nu}_\tau)\nu_\tau$ (red, bottom-left to top-right).
  The distributions shown on the left correspond to only $-0.20<M^2_{\rm{miss}}<0.20$~$(\rm{GeV}/c^2)^2$,
  while we show the entire momentum spectra on the right.
}
\label{fig:mcbkgDATAboomom}
\end{figure*}

\begin{figure}[htbp]
  \centering
  \includegraphics[keepaspectratio=true,width=3.1in,angle=0]{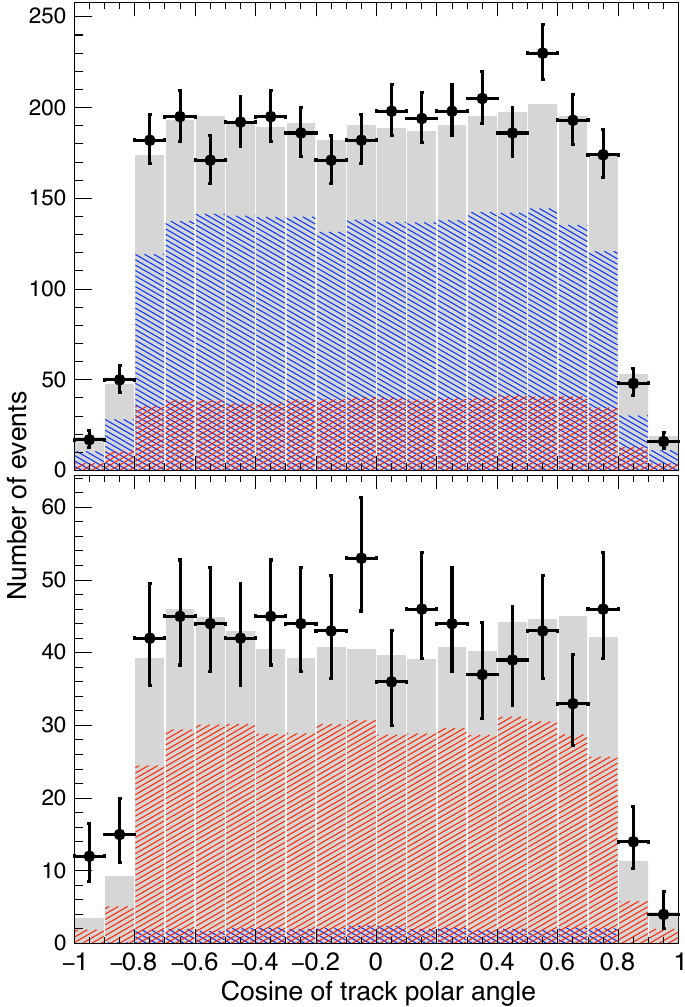}
\caption{Distributions of cosines of polar angles in the $D_s^+$ rest frame of $\mu$-like (top) and $\pi$-like 
  (bottom) signal-candidate tracks, based on the nominal $DT$ selection and the additional requirements
  $1930<M_{\text{inv}}(D^-_s)<1990$~MeV$/c^2$ and $-0.1<M^2_{\rm{miss}}<+0.2$~$(\rm{GeV}/c^2)^2$.
  Black points are data and the overlaid histograms correspond
  to the $40\times$ MC sample scaled to the integrated luminosity of data, with gray shading for the total, 
  and
  crosshatched for the signals, $D^+_s\to\mu^+\nu_\mu$ (blue, top-left to bottom-right)
  and $D^+_s\to\tau^+(\to\pi^+\bar{\nu}_\tau)\nu_\tau$ (red, bottom-left to top-right), respectively.
  The difference between gray and (the sum of red and blue) is the background.
}
\label{fig:mcbkgDATAboocos}
\end{figure}

\subsubsection{\boldmath  Fitting to data}

The fitting procedure described in Sec.~\ref{sec:dtana} accounts for all signal and background processes that have been observed experimentally to date.
We now introduce two additional physics processes that are expected theoretically but
have not yet been observed,
$D_s^+\to\gamma\mu^+\nu_\mu$ and $D_s^+\to\pi^+\pi^0$.

The radiative leptonic decay
$D^+_s\to\gamma\mu^+\nu_\mu$ is not helicity suppressed, but its experimental detection is
difficult and has not yet been achieved.  Past measurements of $D_s^+$ leptonic 
decays~\cite{CLEODs, BES3Ds} have relied on theoretical predictions of the branching fraction for
this mode, with a $1\%$ (relative) systematic uncertainty.  For our analysis we note that the $M^2_{\rm{miss}}$ distribution for 
$D^+_s\to\gamma\mu^+\nu_\mu$ must have a high-side tail and could therefore affect our signal fits
significantly.  We generate a signal MC sample for this process following the procedure of previous BESIII 
studies of $D^+\to\gamma e^+\nu_e$~\cite{BES3Dgamenu} and
$D^+_s\to\gamma e^+\nu_e$~\cite{BES3Dsgamenu}.  This adopts a factorization 
approach~\cite{RadModel} in modeling $D^+_s\to\gamma\mu^+\nu_\mu$ events, 
requiring a minimum photon energy of $10~$MeV.
Figure~\ref{fig:mm2GamMuNu} shows the predicted $M^2_{\rm{miss}}$ distribution for $D^+_s\to\gamma\mu^+\nu_\mu$,
along with distributions for other processes.  Because of the 
high-side tail of the distribution for $D^+_s\to\gamma\mu^+\nu_\mu$, we explicitly include this additional 
PDF in fitting our data.

We estimate the scale of the $D^+_s\to\gamma\mu^+\nu_\mu$ contribution based on theoretical predictions
of the branching fraction.  G. Burdman, J. T. Goldman, and D. Wyler~\cite{radcorr} predict
$\b(D^+_s\to\gamma\mu^+\nu_\mu)\sim10^{-4}$ and 
$\b(D^+\to\gamma\mu^+\nu_\mu)\sim10^{-5}$, while J.~C. Yang and M.~Z. Yang estimate
$\b(D^+\to\gamma\mu^+\nu_\mu)$ to be $3.64\times10^{-5}$~\cite{RadDmunu}.
Comparing these to the measured branching fractions of nonradiative leptonic decays,
$\b(D^+\to\mu^+\nu_\mu)=(3.74\pm0.17)\times10^{-4}$~\cite{pdg2020} and
$\b(D^+_s\to\mu^+\nu_\mu)=(5.49\pm0.16)\times10^{-3}$~\cite{pdg2020}, we adopt the ratio 
$R_\gamma =
\b(D^+_{(s)}\to\gamma\mu^+\nu_\mu)/\b(D^+_{(s)}\to\mu^+\nu_\mu)=0.1$ as a constraint in our fits.
The estimated number of reconstructed radiative events in our signal sample is about $1\%$ of 
the nonradiative events, which is consistent with previous publications (\cite{CLEODs, BES3Ds}).  In assessing 
the systematic uncertainty associated with this assumption, we vary $R_\gamma$ by $\pm0.1$.

The suppressed hadronic decay $D_s^+\to\pi^+\pi^0$ has not been observed, with only an upper limit on 
the branching fraction of $\b(D^+_s\to\pi^+\pi^0)<3.4\times10^{-4}$ at $90\%$ confidence level~\cite{pdg2020}.
Events of this type exhibit a clear peak in $M^2_{\rm{miss}}$, as is demonstrated by MC simulation and shown in 
Fig.~\ref{fig:mm2GamMuNu}.  We do not include $D_s^+\to\pi^+\pi^0$ in our nominal data fit, 
but introduce it with a branching fraction equal to the upper limit as a systematic variation.

\subsubsection{Branching fraction results}

Figure~\ref{fig:data_4180_both} shows our nominal fit to the $4180$ data sample with scaled MC 
components overlaid.
(Figures showing fit results for the other five datasets are provided as supplemental material~\cite{suppmatl}.)
The fit to all data samples yields $2198\pm55$ $D^+_s\to\mu^+\nu_\mu$ events and
$946^{+46}_{-45}$ $D^+_s\to\tau^+(\to\pi^+\bar{\nu}_\tau)\nu_\tau$ events.
The measured branching fractions and statistical uncertainties following from these event yields are 
$\b(D^+_s\to\mu^+\nu_\mu) = (5.35\pm0.13)\times10^{-3}$ and
$\b(D^+_s\to\tau^+\nu_\tau) = (5.21\pm0.25)\%$, 
where we assume
$\b(\tau^+\to\pi^+\bar{\nu}_\tau)=10.82\%$~\cite{pdg2020}.
Our two measurements of the $D_s^+$ leptonic branching fractions
have the best statistical precision to date.

\begin{figure}[htbp]
\centering
\includegraphics[keepaspectratio=true,width=3.1in,angle=0]{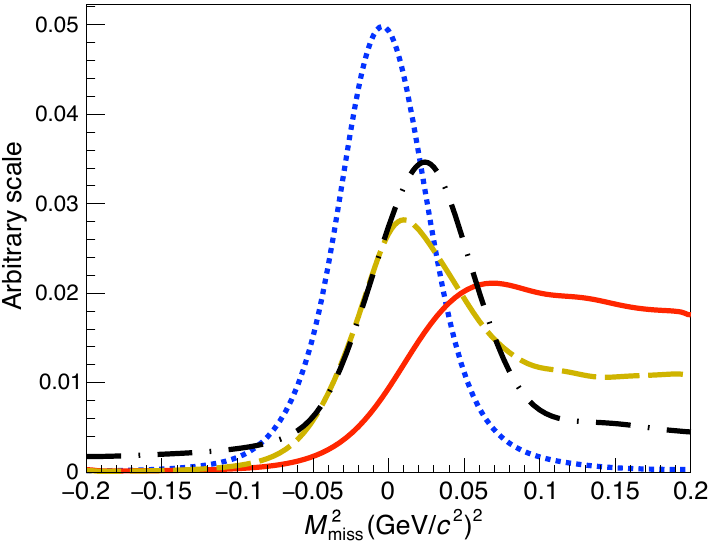}
\caption{Comparison of $M^2_{\rm{miss}}$ distributions for various types of MC
  events for the $\mu$-like case.  (Shapes are very similar for the $\pi$-like case.)
  The dotted-blue line represents $D^+_s\to\mu^+\nu_\mu$, solid-red is
  $D^+_s\to\tau^+(\to\pi^+\bar{\nu}_\tau)\nu_\tau$, dashed-orange is $D^+_s\to\gamma\mu^+\nu_\mu$, 
 and dotted-dashed-black is $D^+_s\to\pi^+\pi^0$. All distributions are normalized to unity.}
\label{fig:mm2GamMuNu}
\end{figure}

\begin{figure*}[htb]
  \centering
  \includegraphics[keepaspectratio=true,width=6.5in,angle=0]{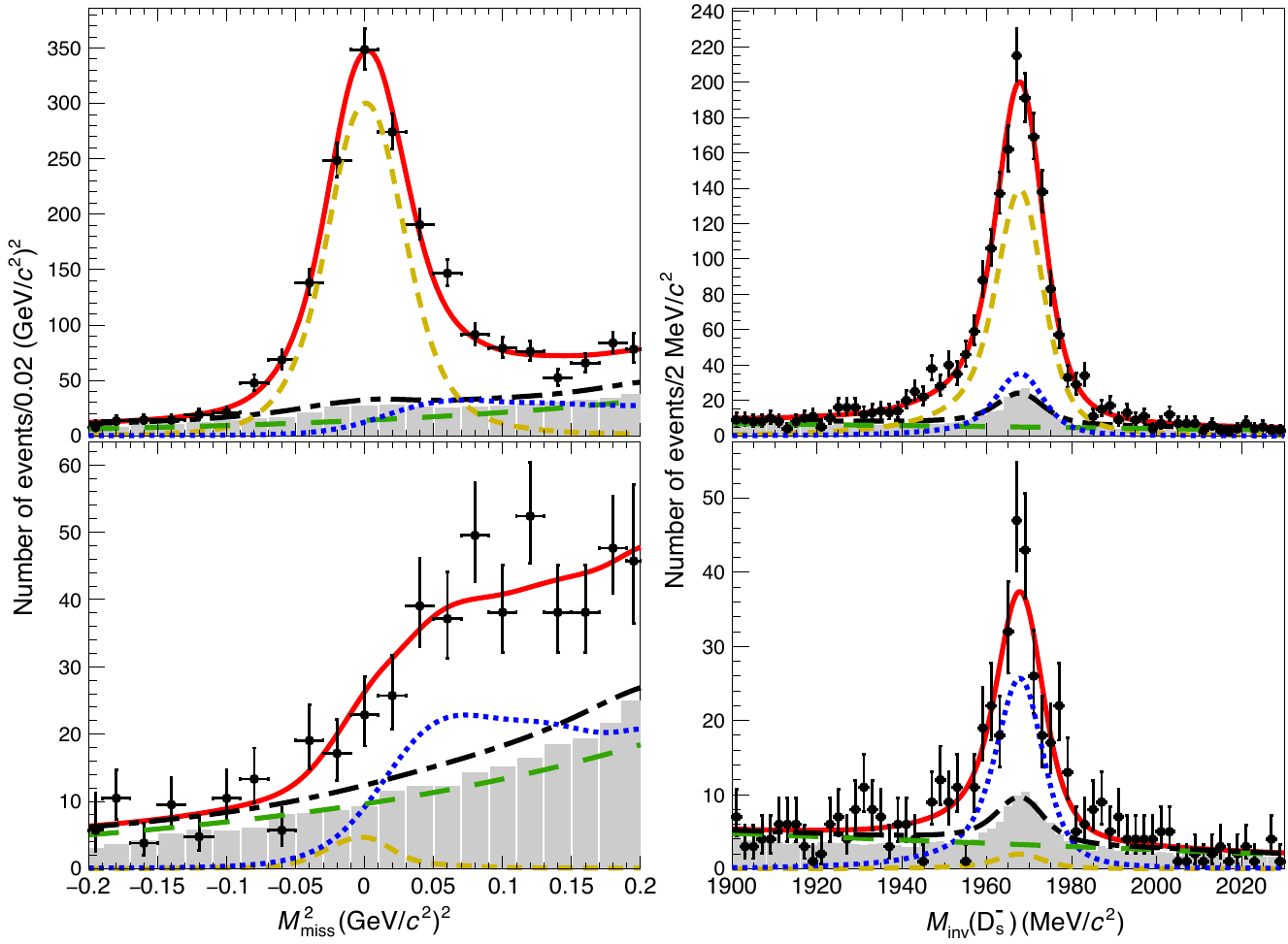}\\
\caption{Projections onto the $M^2_{\rm{miss}}$ (left) and $M_{\text{inv}}(D^-_s)$ (right) axes
  of the two-dimensional fit to 4180 data for the $\mu$-like (top) and the $\pi$-like (bottom) samples.
  Figures showing fit results for the other five datasets are provided as supplemental material~\cite{suppmatl}.
  The black points are data,
the shaded histograms correspond to the $40\times$ background MC sample scaled to the integrated luminosity of data, and the lines
represent the fitted signal and background shapes.
The red-solid, orange-dashed, and blue-dotted lines represent
the total, $D_s^+\to\mu^+\nu_\mu$, and $D_s^+\to\tau^+\nu_\tau$, while
black-dot-dashed and green-long-dashed lines correspond to the total background and the case when both tag and signal sides
are misreconstructed, respectively.}
\label{fig:data_4180_both}
\end{figure*}

We also report the ratio of  the two branching fractions,
$R = \Gamma(D^+_s\to\tau^+\nu_\tau)/\Gamma(D^+_s\to\mu^+\nu_\mu)$
$= 9.73^{+0.61}_{-0.58}$, where the statistical uncertainty includes
the observed anticorrelation between the two components.

An alternative procedure that provides a more statistically precise but model-dependent determination of
$\b(D^+_s\to\tau^+\nu_\tau)$ is to fit our data with the ratio $R$ fixed to the SM 
prediction of $9.75$.  This fit yields a $D^+_s\to\tau^+(\to\pi^+\bar{\nu}_\tau)\nu_\tau$ signal of 
$946\pm18$ events and $\b(D^+_s\to\tau^+\nu_\tau) = (5.22\pm0.10)\%$.

  Because of this anticorrelation between the $D_s^+\to\mu^+\nu_\mu$ and $D_s^+\to\tau^+\nu_\tau$ signal processes,
  caution is necessary in extracting the decay constant $f_{D_s^+}$ and CKM matrix element $|V_{cs}|$
  from an average of the measured branching fractions.  We circumvent this difficulty by requiring lepton flavor universality (LFU),
  which requires that values of $f_{D_s^+}|V_{cs}|$ extracted from $\b(D_s^+\to\mu^+\nu_\mu)$ and $\b(D_s^+\to\tau^+\nu_\tau)$ be identical.
  From Eqs.~\eqref{eq:decayrate} and \eqref{eq:theR}, it can be seen that this LFU constraint is equivalent to the condition $R = 9.75$.
  Thus, in Sec.~\ref{sec:summary} we present our combined average value of $f_{D_s^+}|V_{cs}|$
  by using the measurement obtained with the constraint $R = 9.75$.

\subsubsection{\boldmath {\it CP}-violating  Asymmetries}

We also measure the {\it CP}-violating asymmetries,
\begin{equation}
A_{\it CP} = \frac{\Gamma(D_s^+\to\ell^+\nu_\ell) - \Gamma(D_s^-\to\ell^-\bar{\nu}_\ell)}{
  \Gamma(D_s^+\to\ell^+\nu_\ell) + \Gamma(D_s^-\to\ell^-\bar{\nu}_\ell)},
\end{equation}
\noindent where $\ell = \mu$ or $\tau$. Procedures are identical to those applied to the full sample, 
except that we determine the branching fractions separately for $D_s^+$ and $D_s^-$ .

To search for systematic effects specific to this measurement, we look at
$\Delta N^{\pm}_{tag} = \sum_i\frac{N_{tag,i}^{+}-N_{tag,i}^{-}}{N_{tag,i}^{+}+N_{tag,i}^{-}}$,
where $i$ runs over the thirteen tag modes and
$N_{tag,i}^{\pm} = N^{\pm}_{ST,i}/\epsilon^{\pm}_{ST,i}$ for $D_s^\pm\to i$ mode,
and we combine
the six datasets. We obtain
$\Delta N^{\pm}_{tag} = (+0.6\pm0.8)\%$, consistent
with zero {\it CP} asymmetry, which involves the simulations of charge-dependent tracking
and PID efficiencies. We conservatively assign $1.0\%$
($0.6\%$ and $0.8\%$ combined in quadrature) as a possible
systematic uncertainty due to charge dependence in particle reconstruction
and assign no additional systematic uncertainty to our measurements.  
Potential systematic effects that are associated with our $DT$ fitting procedure are canceled in 
the determination of $A_{\it CP}$.

The nominal fit yields $1123\pm40$ events and $463^{+33}_{-32}$ events
  for $D_s^-\to\mu^-\nu_\mu$ and $D_s^-\to\tau^-\nu_\tau$ candidates, respectively,
  and $1077\pm38$ events and $487^{+33}_{-32}$ events
  for $D_s^+\to\mu^+\nu_\mu$ and $D_s^+\to\tau^+\nu_\tau$ candidates, respectively.
Table~\ref{tab:acp} shows the resultant branching fractions for $D_s^+$ and $D_s^-$,
as well as $A_{\it CP}$, based on the two fitting methods: our principal method, which yields
both $\b(D_s^+\to\mu^+\nu_\mu)$ ($\equiv \b_{\mu\nu}$) and
$\b(D_s^+\to\tau^+\nu_\tau)$  ($\equiv \b_{\tau\nu}$), and the alternative method imposing the SM constraint, 
labeled as $\b_{\tau\nu}^{\text{SM}}$.
The first uncertainties quoted in $A_{\it CP}$ are statistical and the second systematic.
All three $A_{\it CP}$ values show no evidence of {\it CP} violation. This is the first measurement of
$A_{\it CP}(\tau\nu)$ and the most precise determinations to date of both $A_{\it CP}(\mu\nu)$ and 
$A_{\it CP}(\tau\nu)^{\text{SM}}$.

\begin{table}[htb]
\caption{Summary of charge-dependent branching fractions and $A_{\it CP}$ (in $\%$) for $\b(D^+_s\to\mu^+\nu_\mu)$
  and $\b(D^+_s\to\tau^+\nu_\tau)$.
  The uncertainties reported in branching fractions are only statistical.
  The first uncertainties quoted in $A_{\it CP}$ are statistical and the second systematic.
}
\def\1#1#2#3{\multicolumn{#1}{#2}{#3}}
\label{tab:acp}
\begin{center}
\scalebox{1.0}
{
\begin{tabular}{c c c c}
  \hline\hline
  \T\B    & $D_s^-$ & $D_s^+$ & Combined \\
  \hline
  \T\B $\b_{\mu\nu}\times10^3$ & $5.40\pm0.19$ & $5.28\pm0.19$ & $5.35\pm0.13$ \\
  \T\B $\b_{\tau\nu}\times10^2$ & $5.07^{+0.36}_{-0.35}$ & $5.37^{+0.36}_{-0.35}$ & $5.21\pm0.25$ \\
  \T\B $\b_{\tau\nu}^{\text{SM}}\times10^2$ & $5.21\pm0.14$ & $5.20\pm0.14$ & $5.22\pm0.10$ \\
  \hline
  \T\B $A_{\it CP}(\mu^\pm\nu)$  & \multicolumn{3}{c}{$-1.2\pm 2.5\pm 1.0$} \\
  \T\B $A_{\it CP}(\tau^\pm\nu)$  & \multicolumn{3}{c}{$+2.9\pm 4.8\pm 1.0$} \\
  \T\B $A_{\it CP}(\tau^\pm\nu)^{\text{SM}}$  & \multicolumn{3}{c}{$-0.1\pm 1.9\pm 1.0$} \\
  \hline\hline
\end{tabular}
}
\end{center}
\end{table}

\section{SYSTEMATIC UNCERTAINTIES}

We consider a wide variety of potential sources of systematic uncertainty in our measurements of the 
branching fractions $\b(D^+_s\to\mu^+\nu_\mu)$ and $\b(D^+_s\to\tau^+\nu_\tau)$, and their ratio $R$.
Procedures are described in the following two subsections and the resulting estimates are listed in 
Table~\ref{tab:systsummary}.  The sources of systematic uncertainty subdivide into two categories.  
Sources associated with the two-dimensional simultaneous $DT$ fitting procedure affect all measurements, 
while those that are not related to fitting largely cancel in measuring the ratio $R$.

\subsection{\boldmath Nonfitting systematic uncertainties}\label{sec:systnonfit}

We directly estimate the systematic uncertainties associated with the two input branching fractions, 
$\b(D^*_s\to\gamma D_s)$ and $\b(\tau^+\to\pi^+\bar{\nu}_\tau)$, by propagating the uncertainties from 
Ref.~\cite{pdg2020}.

We estimate the systematic uncertainty associated with the reconstruction
of the signal track, $\mu$ or $\pi$, by reconstructing events from the continuum process 
$e^+e^-\to K^+K^-\pi^+\pi^-$.  By comparing the pion reconstruction efficiency over the relevant momentum
range for $4180$ data and MC samples, the reliability of the simulation is found to be better than 1\%.  The stability of tracking 
over our six data samples is demonstrated by consistent performance on control samples of
radiative $\mu$-pair events.  On this basis we assign a $1\%$ systematic uncertainty for our branching fraction
measurements.

The systematic uncertainty associated with the reconstruction of the photon from $D^*_s\to\gamma D_s$
is estimated by reconstructing $J/\psi\to\pi^+\pi^-\pi^0$ events~\cite{BES3GamSyst}.  Comparison of the 
photon-reconstruction efficiency in data and MC samples gives a $1\%$ systematic uncertainty for this source.

In selecting our signal-side sample, we require that there be only one charged track in addition to the 
daughters of the reconstructed tag, as described in Sec.~\ref{sec:dtana}.  We estimate the systematic 
uncertainty for this requirement based on double-hadronic-tag (DHT) events in which a $D_s^-$ is tagged 
in one of our thirteen modes, while the $D_s^+$ decays into either $K^0_S\pi$ or $KK\pi$.  The uncertainty is 
$0.2\%$ for all three branching fraction measurements.

As described in Sec.~\ref{sec:STselect}, we handle events with multiple $ST$ candidates for a given 
$D_s$ mode and charge by choosing the one with $E_\gamma$, the $D^*_s\to\gamma D_s$ photon energy 
in the $D^*_s$ rest frame, closest to the expected value.  We investigate systematic effects in this selection
by comparing efficiencies for DHT events in data and MC samples.  The agreement is found to have some dependence
on event complexity (charged and neutral particle multiplicity),
but is no worse than $1\%$, so we assign this as the systematic  uncertainty
in the best-photon selection for all branching fraction measurements.

Our signal-selection procedure (Sec.~\ref{sec:dtana}) includes three additional requirements that are designed 
to suppress events with unreconstructed particles.  We study systematic uncertainties associated with these
using the same DHT events.  For the requirement $E^\text{max}_\text{neu}<300$~MeV, we compare the
efficiencies
in data and MC samples for the standard
requirement
and probe the stability of the data/MC agreement 
by also testing with
requirements
less or more restrictive
than this by 50~MeV.  We find an uncertainty of $0.3\%$ for all branching fractions.  We similarly probe the $ANG>40\degree$ and $|\cos{\theta_\text{miss}}|<0.90$
efficiencies, although in this case the results with the $D_s$ DHT sample are limited by sample size.
  We augment 
with data collected at $E_{\text{cm}} = 3773$~MeV, with an integrated luminosity of $2.93$~fb$^{-1}$, and
copious production of $\psi(3770)\to D\bar{D}$.  In this sample we measure data and MC
efficiencies for
$D^0$ decays into the three hadronic modes, $K^-\pi^+$, $K^-\pi^+\pi^0$, and $K^-\pi^+\pi^-\pi^+$, and for
the semileptonic decay $\bar{D}^0\to K^+ e^- \bar{\nu}_e$.  Based on these studies, we assign a $1\%$ systematic
uncertainty for the $ANG$ and $|\cos{\theta_\text{miss}}|$ requirements for all branching fraction measurements.


The determination of the $D_s$ leptonic branching fractions with Eqs.~\eqref{eq:dteq} and \eqref{eq:dteqtau} depends 
on the efficiency ratios $\epsilon^{\ell\nu,i}_{DT}/\epsilon^i_{ST}$.  Both $ST$ and $DT$ selection involve reconstructing a 
hadronic $D^-_s$ decay, and we expect the efficiency for this tag reconstruction to depend on the event environment.  The different topologies of leptonic $D_s^+$ decays (only one track) and generic $D_s^+$ 
decays (most with multiple tracks and showers) produce a mode-dependent bias in reconstructing the 
$D_s^-$ tag that may be imperfectly modeled in the MC simulation.  We estimate the systematic uncertainty associated 
with this effect by studying the BESIII detector's tracking and PID efficiencies for events with different particle
multiplicities using the large $E_{\text{cm}} = 3773$~MeV data sample mentioned earlier.  The size 
of this uncertainty varies among the three branching fraction and $R$ measurements, as is shown in 
Table~\ref{tab:systsummary}.

\subsection{\boldmath Fitting systematic uncertainties}\label{sec:systfit}

To assess the systematic uncertainties associated with our fitting
procedure, we generate toy Monte Carlo samples based on
the observed data distributions.
We fit to these toy samples while
varying an analysis selection requirement
(or fitting procedure, PID requirement, etc.)
and take the difference between the averages of these ensembles with the 
nominal fit procedure and with the alternative procedure and assign it as a 
systematic uncertainty.  Table~\ref{tab:systsummary} shows that these estimated 
systematic uncertainties vary significantly among the measurements of the three branching fraction and 
$R$.

The uncertainty in the determination of the denominators 
in Eqs.~\eqref{eq:dteq} and \eqref{eq:dteqtau} arises
mainly from fitting to $M_{\text{inv}}(D^-_s)$ for $ST$ candidates. 
The dominant effect comes from
background and signal shapes (including the convolved 
Gaussian functions, which are independently determined for $4180$, $4190-4220$ and $4230$).
We also investigate the contamination from $e^+e^-\to\gamma_{ISR}D^+_sD^-_s$, and find the uncertainty 
associated with this to be negligible.

For a conservative estimate of the uncertainty due to the assumption of a fixed ratio 
$R_\gamma = \b(D^+_{(s)}\to\gamma\mu^+\nu_\mu)/\b(D^+_{(s)}\to\mu^+\nu_\mu)=0.1$, 
we vary $R_\gamma$ by $\pm0.1$.

To allow for the possible effect of the unobserved mode $D^+_s\to\pi^+\pi^0$, which is excluded from 
our nominal fit, we include the PDF for this mode in an alternative fit, with the normalization set to the 
experimental upper limit, $\b(D^+_s\to\pi^+\pi^0)<3.4\times10^{-4}$.

We consider a possible systematic uncertainty due to $\pi$-ID efficiency,
measuring the effect with the $D_s$ and $D^0$ data samples mentioned 
in Sec.~\ref{sec:dtana}.
The uncertainty due to the rate for misidentification of $\mu$ as $\pi$ is estimated by comparing
the rate between data and our MC simulation in the $D_s^+\to\mu^+\nu_\mu$ events, in which
we heavily suppress the contribution from  $D_s^+\to\tau^+(\tau^+\to\pi^+\bar{\nu}_\tau)\nu_\tau$
by requiring the signal track to penetrate deep into our MUC, as is done in Ref.~\cite{bes3dsmunu}.

In the nominal fitting procedure, we share two convolved Gaussian functions
[one each for $M^2_{\rm{miss}}$ and $M_{\text{inv}}(D_s^-)$] over the six data samples, 
effectively assuming that any data/MC differences are independent of 
$E_{\text{cm}}$ and changes in running conditions.
We estimate a possible uncertainty due to this assumption with an alternative 
fit using independent Gaussian functions for each of the $4180$, $4190-4220$ and $4230$ data groups.

The relative size of the background component arising from misreconstruction on the tag side 
is fixed according to MC simulation in our nominal fit procedure.
We estimate the potential systematic uncertainty introduced by this constraint with an alternative 
fit allowing this component to float freely. 

The relative size of the wrong-photon background component
is also fixed in our nominal fitting procedure.
For a systematic test, we vary this by $\pm 1.4\%$ (relative), the quadrature sum of 
a  $1.0\%$ uncertainty for photon reconstruction and a second $1.0\%$ uncertainty for
the best-photon selection method.

The size of the background component in which the signal track is misreconstructed is fixed in our nominal procedure to MC simulation, normalized to the $ST$ yields in data. The primary source is the decay 
$D^+_s\to K^0\pi^+$, and we use the the uncertainties in $\b(D^+_s\to K^0_S\pi^+)$~\cite{pdg2020} and 
in our $N_{ST}$ determination to estimate the systematic uncertainty in the estimate of this background 
component.

Background events in which both the tag side and the signal track
are misreconstructed are parametrized in the nominal fit with a first-order Chebyshev polynominal for $M_{\text{inv}}(D^-_s)$ and a first-order exponential polynomial for $M^2_{\rm{miss}}$.
We use a MC-based shape for $M_{\text{inv}}(D^-_s)$ and a first-order Chebyshev polynominal for $M^2_{\rm{miss}}$ in
an alternative fit to estimate a possible systematic uncertainty due to the assumed background shape.

MC studies show that $D^+_s\to\tau^+\nu_\tau$ events with $\tau^+$ decays into
final states other than $\pi^+\bar{\nu}_\tau$
that are counted as signal are dominated by $\tau^+\to\mu^+\nu_\mu\bar{\nu}_\tau$ and 
$\tau^+\to\pi^+\pi^0\bar{\nu}_\tau$ for the $\mu$-like sample and by $\tau^+\to\pi^+\pi^0\bar{\nu}_\tau$ 
and $\tau^+\to e^+\nu_e\bar{\nu}_\tau$ for the $\pi$-like sample.
We estimate the uncertainty in the estimate 
of these events with variations based on the
uncertainties in the measured branching fractions~\cite{pdg2020}.

In our nominal fitting procedure, we fix the relative yields of signal
between the  $\mu$-like and $\pi$-like samples according to MC simulation.
The $\mu$-like and $\pi$-like samples are defined by
$E_{\rm{EMC}} \le 300$~MeV and $E_{\rm{EMC}} > 300$~MeV, respectively.
Thus, to assess the systematic uncertainty associated with this criterion,
we look at distributions  of $E_{\rm{EMC}}$ and see how well our MC
agrees with data.  We look at distributions for muons from $D_s^+\to\mu^+\nu_\mu$
and for pions from $D^{*+}D^-$ and $D^{*+}D^{*-}$, with $D^{*+}\to D^0\pi^+$ and 
$D^0\to K^-\pi^+$, where the $\pi^+$
coming from the $D^0$ decay deposits $E_{\rm{EMC}}$, while requiring
$800<|\vec{p}_\pi|<1100$~MeV$/c$ to match our signal pion and muon tracks.
We observe a $4\%$ (relative) difference in partitioning rates between data and MC samples.
We vary the rate by $\pm4\%$ to estimate this systematic uncertainty.

As an alternative fitting procedure, we constrain the yields of $D_s^+\to\mu^+\nu_\mu$ and
  $D_s^+\to\tau^+\nu_\tau$ to the SM expectation for the ratio $R = 9.75$, derived from Eq.~\eqref{eq:theR}.
  The uncertainty in this prediction arises from the input particle masses, which are precisely known.
We estimate a possible systematic uncertainty due to this constraint
by varying $R$ by $\pm0.01$.

\begin{table}[htb]
  \caption{Systematic uncertainties on
    $\b_{\mu\nu}$, $\b_{\tau\nu}$, $\b_{\tau\nu}^{\text{SM}}$, and $R$.
   The notation ``cncl.'' indicates that a systematic uncertainty cancels in the calculation 
   of the branching fraction ratio $R$, and ``neg.'' signifies that an uncertainty is negligible.
}
\label{tab:systsummary}
\begin{center}
\scalebox{1.0}
{
  \begin{tabular}{C{3.9cm} C{0.9cm} C{0.9cm} C{0.9cm} C{0.9cm}}
\hline\hline
 Rel. Syst. Uncertainty ($\%$) &  $\b_{\mu\nu}$ &  $\b_{\tau\nu}$ & $\b_{\tau\nu}^{\text{SM}}$ & $R$ \\
\hline
$\Delta\b(D_s^*\to\gamma D_s)$ & $0.7$ & $0.7$ & $0.7$ & cncl. \\
$\Delta\b(\tau^+\to\pi^+\bar{\nu}_\tau)$ & $\cdot\cdot\cdot$ & $0.5$ & $0.5$ & $0.5$ \\
$\mu$ or $\pi$ tracking & $1.0$ & $1.0$  & $1.0$ & cncl. \\
Photon reconstruction & $1.0$ & $1.0$ & $1.0$ & cncl. \\
$N_{tk} = 1$ & $0.2$ & $0.2$  & $0.2$ & cncl. \\
$E^\text{max}_\text{neu} < 300$~MeV & $0.3$ &  $0.3$  &  $0.3$ & cncl.  \\
Best photon selection & $1.0$ & $1.0$ & $1.0$ & cncl. \\
$ANG$ and $|\cos{\theta_\text{miss}}|$ & $1.0$ & $1.0$  & $1.0$ & cncl. \\
Tag bias & $0.4$ & $0.3$ & $0.4$ & $0.1$  \\
  \hline
$N_{ST}$ based norm. & $0.8$ & $0.8$ & $0.8$ & $0.1$ \\
$\Delta\Gamma(D^+_s\to\gamma\mu^+\nu_\mu)$ & $0.6$ & $1.0$ & $0.7$ & $0.4$ \\
$\Delta\b(D^+_s\to\pi^+\pi^0)$ & $0.1$ & $0.3$ & $0.1$ & $0.2$\\
$\pi$-ID & $0.6$ & $0.5$ & $0.6$ & $1.1$\\
Signal shape & $0.2$ & $0.5$ & $0.2$ & $0.6$ \\
Wrong tag & $1.1$ & $1.4$  & $0.3$ & $2.4$ \\
Wrong photon & $0.1$ & $0.2$ & $0.1$ & $0.1$ \\
Wrong $\mu$ or $\pi$ & neg. & $0.2$ & $0.1$ & $0.2$ \\
Wrong both tag and trk & $1.0$ & $0.8$ & $0.6$ & $1.8$ \\
Other $\tau$ decays & neg. & neg.  & neg. & neg. \\
$\mu/\pi$-like separation & $0.3$ & $0.6$ & $0.6$ & $1.5$\\
$\Delta R$ & $\cdot\cdot\cdot$ & $\cdot\cdot\cdot$ & $0.1$ & $\cdot\cdot\cdot$\\
\hline
Total syst. uncertainty & $2.9$ & $3.2$ & $2.7$ & $3.7$\\
  \hline\hline
\end{tabular}
}
\end{center}
\end{table}

\section{SUMMARY and comparisons to other experimental results}\label{sec:summary}

Our measurements are summarized in Table~\ref{tab:resultsummary},
along with previously published experimental results.
In this section, the first uncertainty quoted is statistical
and the second is systematic.
We measure the absolute branching fraction
$\b(D^+_s\to\tau^+\nu_\tau) = (5.21\pm0.25\pm0.17)\%$,
which is the most precise measurement to date and is in
agreement with the SM prediction of 
$\b(D^+_s\to\tau^+\nu_\tau) = (5.221\pm0.018)\%$ (see Sec.~\ref{sec:intro} for the predicted branching fractions).
For the ratio of the two decay widths,
we obtain $R = \Gamma(D^+_s\to\tau^+\nu_\tau)/\Gamma(D^+_s\to\mu^+\nu_\mu) =
9.73^{+0.61}_{-0.58}\pm0.36$, which is also consistent with
the SM prediction of $9.75$. The precision of these
measurements is limited by the sample size.

By constraining the ratio of the yields of $D^+_s\to\mu^+\nu_\mu$ and $D^+_s\to\tau^+\nu_\tau$
to the SM expectation $R = 9.75$, we gain statistical sensitivity
and obtain
$\b(D^+_s\to\tau^+\nu_\tau) = (5.22\pm0.10\pm0.14)\%$,
which is limited by its systematic uncertainty.

We also obtain $\b(D^+_s\to\mu^+\nu_\mu) =
(5.35\pm0.13\pm0.16)\times10^{-3}$, which is
again the most precise to date. It is consistent
with the previously published BESIII result of 
$(5.49\pm0.16\pm0.15)\times10^{-3}$~\cite{bes3dsmunu}, which only analyzed the 4180 data, and
 supersedes that result. 
Note that the analysis methods and background compositions for these two BESIII analyses
are very different. In Ref. ~\cite{bes3dsmunu}, we require $\mu$ identification with the MUC 
subdetector to suppress $\pi$ from $D_s^+\to\tau^+\nu_\tau$, resulting in a smaller systematic uncertainty.

We also measure the {\it CP}-violating asymmetries
$A_{\it CP}(\mu^\pm\nu) = (-1.2\pm2.5\pm1.0)\%$ and
$A_{\it CP}(\tau^\pm\nu) = (+2.9\pm4.8\pm1.0)\%$.
The former is the most precise to date and the latter is a first measurement.
With the SM constraint, we have $A_{\it CP}(\tau^\pm\nu) = (-0.1\pm1.0\pm1.0)\%$,
which is also the most precise to date. All three $A_{\it CP}$ results show no evidence of {\it CP} violation.

The measured $\b(D_s^+\to\tau^+\nu_\tau)$ with the SM constraint leads to
\[
f_{D^+_s}|V_{cs}| = (243.2\pm2.3\pm3.3\pm1.0)~\text{MeV},
\]
where the last uncertainty comes from the external inputs (lepton masses, $m_{D_s}$, and the 
$D_s^+$ lifetime~\cite{pdg2020}).  Table~\ref{tab:resultsummary} presents results based on other 
fitting schemes.  By taking $|V_{cs}| = 0.97320\pm0.00011$ from the constrained global fit~\cite{pdg2020}
and the average of the recent four-flavor LQCD predictions,  
$f_{D^+_s} = (249.9\pm0.5$)~MeV~\cite{flag2019},
one finds an expected product of these of $(243.2\pm0.5)$~MeV, in excellent agreement with our result.

By taking $|V_{cs}| = 0.97320\pm0.00011$~\cite{pdg2020}
as an input, we obtain
\begin{equation*}
  \begin{array}{r c l c}
    f_{D^+_s} & = & 249.8\pm3.0\pm3.7\pm1.0~\text{MeV}, &  \\
              &   & 249.7\pm6.0\pm4.1\pm1.0~\text{MeV}, & \text{and} \\
        &   & 249.9\pm2.4\pm3.4\pm1.0~\text{MeV}, \\
\end{array}
\end{equation*}
\noindent from $\b_{\mu\nu}$, $\b_{\tau\nu}$, and $\b_{\tau\nu}$ with the SM constraint, respectively.
All of these are in excellent agreement with the LQCD predictions. 
Similarly, by taking $f_{D^+_s} = 249.9\pm0.5$~MeV~\cite{flag2019}
as an input, we arrive at
\begin{equation*}
  \begin{array}{r c l c}
    |V_{cs}| & = & 0.973\pm0.012\pm0.015\pm0.004, & \\
             & & 0.972\pm0.023\pm0.016\pm0.004, & \text{and} \\
      & & 0.973\pm0.009\pm0.013\pm0.004, & \\
  \end{array}
\end{equation*}
\noindent based on $\b_{\mu\nu}$, $\b_{\tau\nu}$, and $\b_{\tau\nu}$ with the SM constraint, respectively.
Again they are in agreement with
the global fit result~\cite{pdg2020}.

  Note that in the extraction of $f_{D_s^+}$  and comparison to the global fit result above,
  we treat as negligible the correlation arising from the overlap of $\sim50\%$ of our data sample with that of Ref. ~\cite{bes3dsmunu},
  which is incorporated in the global fit result~\cite{pdg2020}.

Table~\ref{tab:resultsummary}
also shows averages of $f_{D_s^+}|V_{cs}|$ for the current work and previous measurements.
We combine the results from the two modes, $\mu^+\nu$ and $\tau^+\nu$, under the assumption of
the LFU, employing those measurements with the constraint $R = 9.75$
to obtain the combined average value.
As the analysis technique utilized in Refs.~\cite{BES34009, CLEOpi}
  is essentially identical to this work, we also employ their measurements with the SM constraint for the calculation
of the combined average.

  We can also use our $D_s^+$ and previous $D^+$ results to compute 
  $f_{D^+_s}/f_{D^+}$.  LQCD predicts this with great precision, specifically 
  $f_{D^+_s}/f_{D^+} = 1.1783\pm0.0016$ in Ref.~\cite{flag2019}.
  From $\b(D^+\to\mu^+\nu_\mu) = (3.74\pm0.17)\times10^{-4}$~\cite{pdg2020},
  $\tau_{D^+} = 1.040(7)$~ps ~\cite{pdg2020}, and $|V_{cd}| = 0.22636\pm0.00048$~\cite{pdg2020},
  we have $f_{D^+} = 202.9\pm4.7$~MeV. Combining this with the $f_{D_s^+}$ we
  obtain with the SM constraint, we find $f_{D^+_s}/f_{D^+} = 1.232\pm0.035$, consistent with
  the LQCD prediction within $1.5\sigma$. The uncertainty on this ratio is now driven by
  the statistical uncertainty on the measurement of $\b(D^+\to\mu^+\nu_\mu)$.

\begin{table*}[htb]
  \caption{Our measured branching fractions and their corresponding
    products of the decay constant $f_{D^+_s}$ and the CKM matrix-element magnitude $|V_{cs}|$.
    The reconstructed $\tau$-decay mode for each measurement is indicated in parentheses.
    Other experimental results are also shown for comparison.
    The entry labeled ``BESIII @ $4180$'' is the BESIII result from Ref.~\cite{bes3dsmunu},
    based only on the $4180$ data sample and requiring positive muon identification with the MUC.
    ``Averages''
    are obtained by weighting
    both statistical and systematic uncertainties, but
    not the third uncertainties, which are dominated by the uncertainty of the $D_s$ lifetime.
    ``LQCD + PDG'' represents an expected value, as is explained in the text.
}
\label{tab:resultsummary}
\begin{center}
  \def\1#1#2#3{\multicolumn{#1}{#2}{#3}}
\scalebox{1.0}
{
  \begin{tabular}{L{4.0cm} C{3cm} C{4.0cm} C{4.0cm}}
\hline\hline
    Experiment &  Mode & $\b(\%)$ & $f_{D_s^+}|V_{cs}|$ (MeV)\\
    \hline
    {\bf This work}       & $\pmb{\tau^+\nu(\pi^+\bar{\nu})}$  & $\pmb{5.21\pm0.25\pm0.17}$ & $\pmb{243.0\pm5.8\pm4.0\pm1.0}$ \\
    {\bf This work $\pmb{(R=9.75)}$} & $\pmb{\tau^+\nu(\pi^+\bar{\nu})}$ & $\pmb{5.22\pm0.10\pm0.14}$ & $\pmb{243.2\pm2.3\pm3.3\pm1.0}$\\
    BESIII~\cite{BES34009} & $\tau^+\nu(\pi^+\bar{\nu})$   & $3.28\pm1.83\pm0.37$ & $192.8\pm44.2\pm10.9\pm0.8$ \\
    BESIII $(R=9.75)$~\cite{BES34009} & $\tau^+\nu(\pi^+\bar{\nu})$   & $4.83\pm0.65\pm0.26$ & $233.9\pm15.9\pm5.1\pm0.9$ \\
    CLEO~\cite{CLEOel}      & $\tau^+\nu(e^+\nu\bar{\nu})$ & $5.30\pm0.47\pm0.22$ & $245.1\pm10.9\pm5.1\pm1.0$ \\
    CLEO~\cite{CLEOpi}     & $\tau^+\nu(\pi^+\bar{\nu})$   & $6.42\pm0.81\pm0.18$ & $269.7\pm17.2\pm3.8\pm1.1$ \\
    CLEO $(R=9.75)$~\cite{CLEOpi}     & $\tau^+\nu(\pi^+\bar{\nu})$   & $5.77\pm0.36\pm0.18$ & $255.7\pm8.0\pm4.0\pm1.0$ \\
    CLEO~\cite{CLEOro}     & $\tau^+\nu(\rho^+\bar{\nu})$ & $5.52\pm0.57\pm0.21$ & $250.1\pm13.0\pm4.8\pm1.0$ \\
    BaBar~\cite{BaBarlnu}    & $\tau^+\nu(e^+(\mu^+)\nu\bar{\nu})$ & $4.96\pm0.37\pm0.57$ & $237.1\pm8.9\pm13.7\pm1.0$ \\
    Belle~\cite{Bellelnu}       & $\tau^+\nu(\pi^+\bar{\nu}, e^+(\mu^+)\nu\bar{\nu})$ & $5.70\pm0.21^{+0.31}_{-0.30}$ & $254.1\pm4.7\pm7.0\pm1.0$\\
    \hline
    Average$^{\text{a}}$        & $\tau^+\nu$                            & $5.40\pm0.19$  & $247.4\pm4.3\pm1.0$ \\
    \hline
    {\bf This work}     & $\pmb{\mu^+\nu}$ & $\pmb{0.535\pm0.013\pm0.016}$ & $\pmb{243.1\pm3.0\pm3.6\pm1.0}$ \\
    BESIII @ $4180$~\cite{bes3dsmunu} & $\mu^+\nu$ & $0.549\pm0.016\pm0.015$ & $246.2\pm3.6\pm3.4\pm1.0$ \\
    BESIII~\cite{BES34009} & $\mu^+\nu$ & $0.517\pm0.075\pm0.021$ & $238.9\pm17.5\pm4.9\pm0.9$ \\
    CLEO~\cite{CLEOpi}      & $\mu^+\nu$ & $0.565\pm0.045\pm0.017$ & $249.8\pm10.0\pm3.8\pm1.0$ \\
    BaBar~\cite{BaBarlnu}    & $\mu^+\nu$ & $0.602\pm0.038\pm0.034$ & $257.8\pm8.2\pm7.3\pm1.0$ \\
    Belle~\cite{Bellelnu}       & $\mu^+\nu$ & $0.531\pm0.028\pm0.020$ & $242.2\pm6.4\pm4.6\pm1.0$ \\
    \hline
    Average$^{\text{b}}$      & $\mu^+\nu$ & $0.543\pm0.015$ & $244.8\pm3.5\pm1.0$\\
    \hline
    Average$^{\text{c}}$      & $\mu^+\nu + \tau^+\nu$ & $\cdot\cdot\cdot$ &  $246.1\pm2.8\pm1.0$\\
    \hline\hline
    LQCD + PDG &  $\cdot\cdot\cdot$ & $\cdot\cdot\cdot$ & $243.2\pm0.5$ \\
    \hline\hline
    \1{4}{l}{$^{\text{a}}$It excludes ``{\bf This work $\pmb{(R=9.75)}$}'', ``BESIII $(R=9.75)$~\cite{BES34009}'', and ``CLEO $(R=9.75)$~\cite{CLEOpi}''.} \\
    \1{4}{l}{$^{\text{b}}$It excludes ``BESIII @ $4180$~\cite{bes3dsmunu}''.} \\
    \1{4}{l}{$^{\text{c}}$It excludes ``{\bf This work}'', ``BESIII @ $4180$~\cite{bes3dsmunu}'', ``BESIII~\cite{BES34009}'', and ``CLEO~\cite{CLEOpi}''.} \\
\end{tabular}
}
\end{center}
\end{table*}

\acknowledgments
The BESIII Collaboration thanks the staff of BEPCII and the computing center for their hard efforts. This work is supported in part by National Key Research and Development Program of China under Contracts Nos. 2020YFA0406400, 2020YFA0406300; National Natural Science Foundation of China (NSFC) under Contracts Nos. 11625523, 11635010, 11735014, 11822506, 11835012, 11935015, 11935016, 11935018, 11961141012; the Chinese Academy of Sciences (CAS) Large-Scale Scientific Facility Program; Joint Large-Scale Scientific Facility Funds of the NSFC and CAS under Contracts Nos. U1732263, U1832207; CAS Key Research Program of Frontier Sciences under Contracts Nos. QYZDJ-SSW-SLH003, QYZDJ-SSW-SLH040; 100 Talents Program of CAS; INPAC and Shanghai Key Laboratory for Particle Physics and Cosmology; ERC under Contract No. 758462; European Union Horizon 2020 research and innovation programme under Contract No. Marie Sklodowska-Curie grant agreement No 894790; German Research Foundation DFG under Contracts Nos. 443159800, Collaborative Research Center CRC 1044, FOR 2359, FOR 2359, GRK 214; Istituto Nazionale di Fisica Nucleare, Italy; Ministry of Development of Turkey under Contract No. DPT2006K-120470; National Science and Technology fund; Olle Engkvist Foundation under Contract No. 200-0605; STFC (United Kingdom); The Knut and Alice Wallenberg Foundation (Sweden) under Contract No. 2016.0157; The Royal Society, UK under Contracts Nos. DH140054, DH160214; The Swedish Research Council; U. S. Department of Energy under Contracts Nos. DE-FG02-05ER41374, DE-SC-0012069.
%


\pagebreak
\widetext
\clearpage
~\vspace{2cm}
\begin{center}
\textbf{\large Supplemental Material}
\end{center}
\setcounter{equation}{0}
\setcounter{table}{0}
\setcounter{page}{1}
\makeatletter

\begin{center}
\parbox{14cm}{Figures~\ref{fig:data_both_mu_mm2}-\ref{fig:data_both_pi_inv} show our nominal fits to the all six data sets. \label{sec:supp}}
\end{center}
~\vspace{1cm}

\begin{figure*}[htbp]
  \centering
  \includegraphics[keepaspectratio=true,width=7.0in,angle=0]{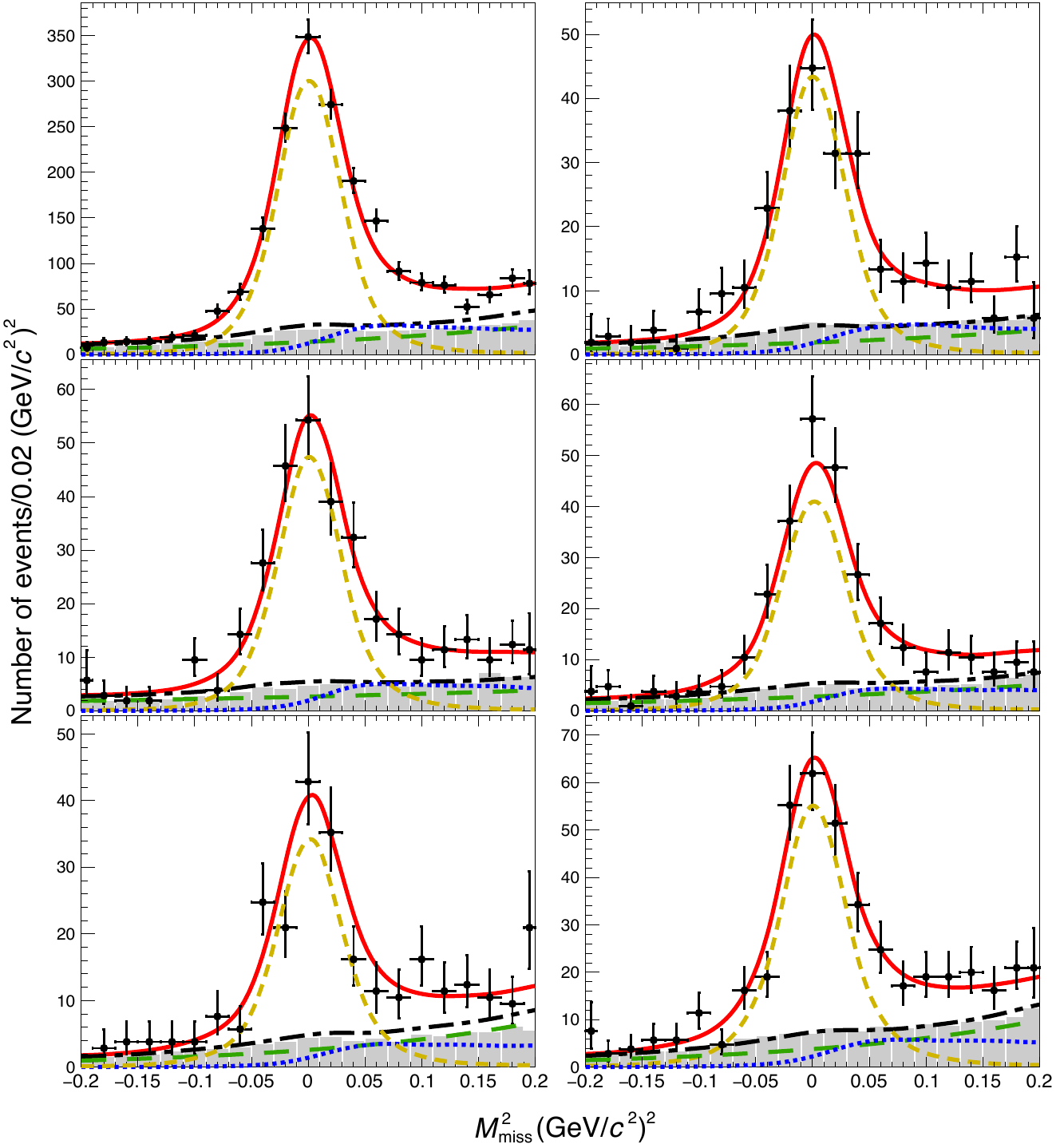}\\
\caption{Projections onto the $M^2_{\rm{miss}}$ axis
  of the two-dimensional fit to $4180$ data (top left), $4190$ data (top right),
  $4200$ data (middle left), $4210$ data (middle right), $4220$ data (bottom left), and
  $4230$ data (bottom right)
  for the $\mu$-like sample.
  The black points are data,
the shaded histograms correspond to the $40\times$ background MC sample scaled to the integrated luminosity of data, and the lines
represent the fitted signal and background shapes.
The red-solid, orange-dashed, and blue-dotted lines represent
the total, $D_s^+\to\mu^+\nu_\mu$, and $D_s^+\to\tau^+\nu_\tau$, while
black-dot-dashed and green-long-dashed lines correspond to the total background and the case when both tag and signal sides
are misreconstructed, respectively.
}
\label{fig:data_both_mu_mm2}
\end{figure*}

\begin{figure*}[htbp]
  \centering
  \includegraphics[keepaspectratio=true,width=7.0in,angle=0]{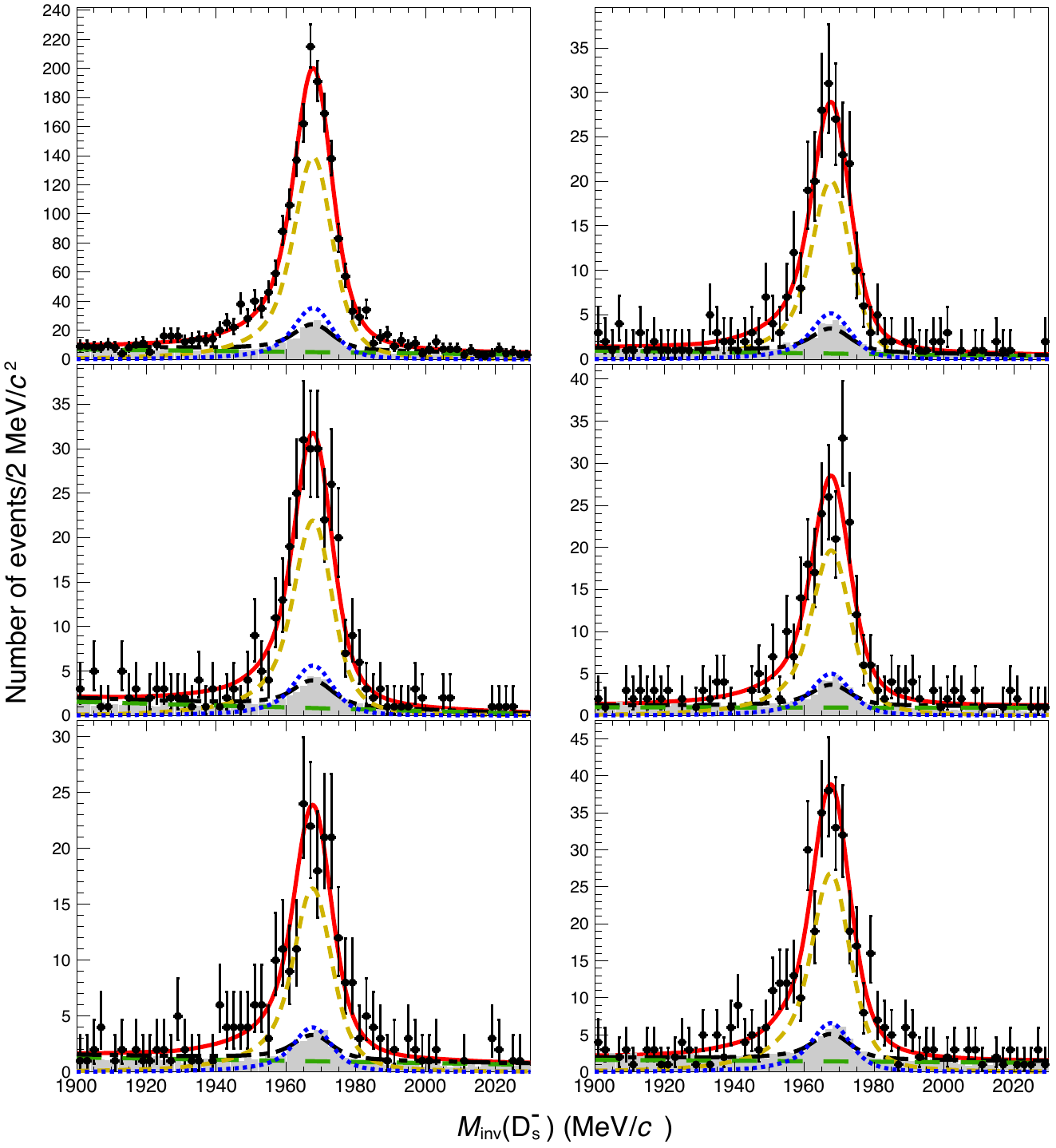}\\
  \caption{Projections onto the $M_{\text{inv}}(D^-_s)$ axis
  of the two-dimensional fit to $4180$ data (top left), $4190$ data (top right),
  $4200$ data (middle left), $4210$ data (middle right), $4220$ data (bottom left), and
  $4230$ data (bottom right)
  for the $\mu$-like sample.
  The black points are data,
the shaded histograms correspond to the $40\times$ background MC sample scaled to the integrated luminosity of data, and the lines
represent the fitted signal and background shapes.
The red-solid, orange-dashed, and blue-dotted lines represent
the total, $D_s^+\to\mu^+\nu_\mu$, and $D_s^+\to\tau^+\nu_\tau$, while
black-dot-dashed and green-long-dashed lines correspond to the total background and the case when both tag and signal sides
are misreconstructed, respectively.
}
\label{fig:data_both_mu_inv}
\end{figure*}

\begin{figure*}[htbp]
  \centering
  \includegraphics[keepaspectratio=true,width=7.0in,angle=0]{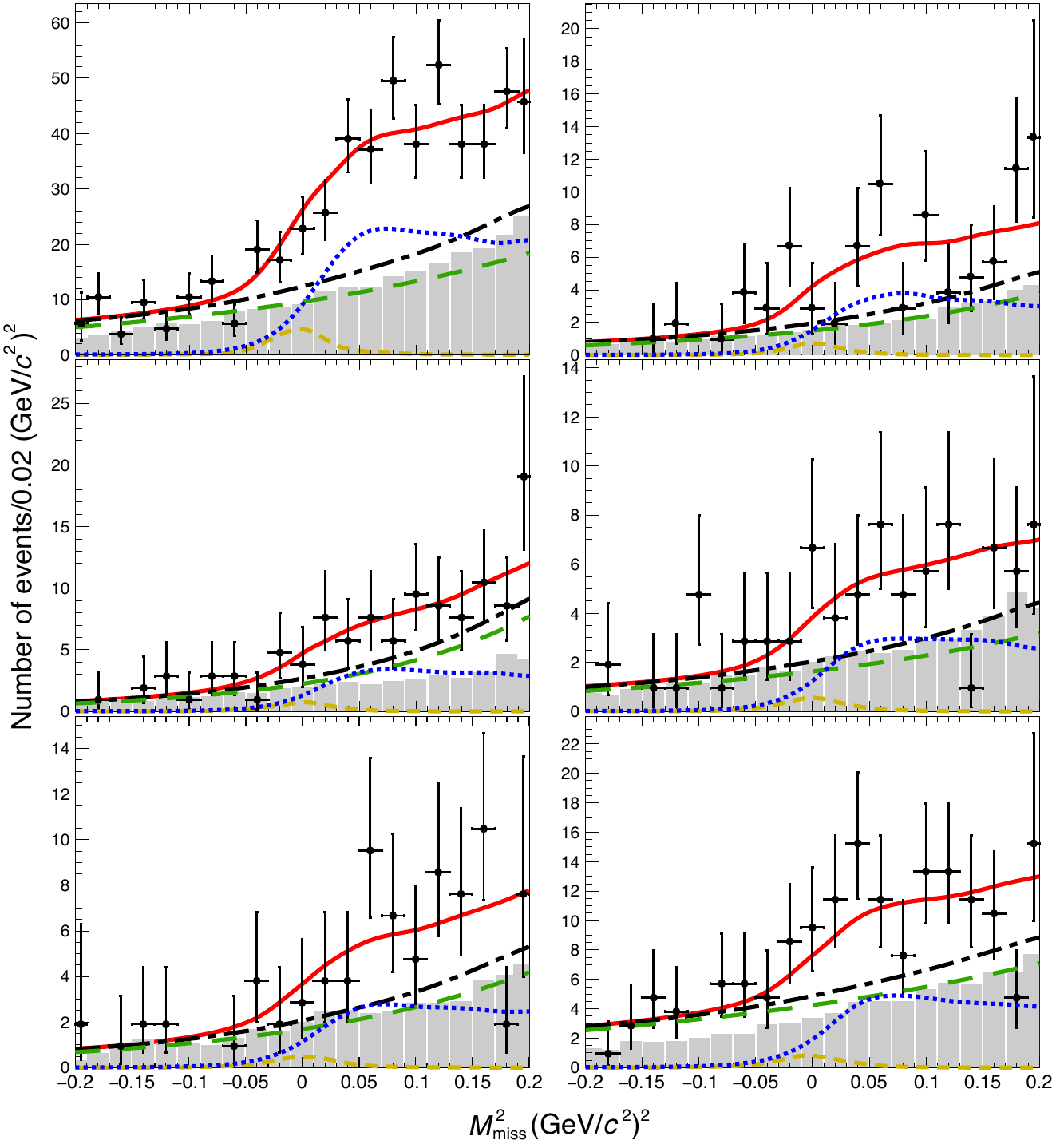}\\
  \caption{Projections onto the $M^2_{\rm{miss}}$ axis
  of the two-dimensional fit to $4180$ data (top left), $4190$ data (top right),
  $4200$ data (middle left), $4210$ data (middle right), $4220$ data (bottom left), and
  $4230$ data (bottom right)
  for the $\pi$-like sample.
  The black points are data,
the shaded histograms correspond to the $40\times$ background MC sample scaled to the integrated luminosity of data, and the lines
represent the fitted signal and background shapes.
The red-solid, orange-dashed, and blue-dotted lines represent
the total, $D_s^+\to\mu^+\nu_\mu$, and $D_s^+\to\tau^+\nu_\tau$, while
black-dot-dashed and green-long-dashed lines correspond to the total background and the case when both tag and signal sides
are misreconstructed, respectively.
}
\label{fig:data_both_pi_mm2}
\end{figure*}

\begin{figure*}[htbp]
  \centering
  \includegraphics[keepaspectratio=true,width=7.0in,angle=0]{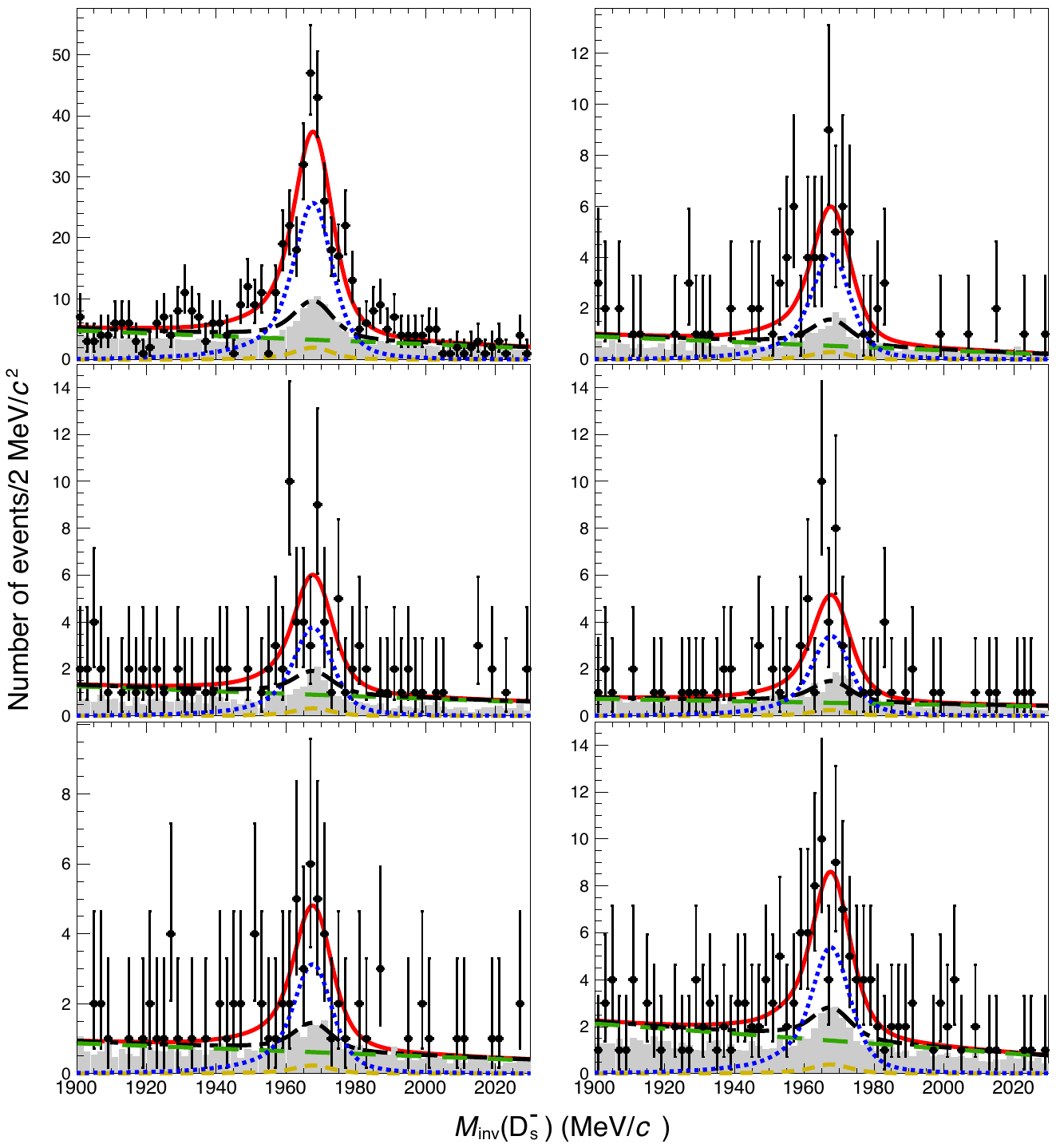}\\
  \caption{Projections onto the $M_{\text{inv}}(D^-_s)$ axis
  of the two-dimensional fit to $4180$ data (top left), $4190$ data (top right),
  $4200$ data (middle left), $4210$ data (middle right), $4220$ data (bottom left), and
  $4230$ data (bottom right)
  for the $\pi$-like sample.
  The black points are data,
the shaded histograms correspond to the $40\times$ background MC sample scaled to the integrated luminosity of data, and the lines
represent the fitted signal and background shapes.
The red-solid, orange-dashed, and blue-dotted lines represent
the total, $D_s^+\to\mu^+\nu_\mu$, and $D_s^+\to\tau^+\nu_\tau$, while
black-dot-dashed and green-long-dashed lines correspond to the total background and the case when both tag and signal sides
are misreconstructed, respectively.
}
\label{fig:data_both_pi_inv}
\end{figure*}

\end{document}